\documentclass[12pt,a4paper,final]{iopart}

\usepackage{iopams}  
\expandafter\let\csname equation*\endcsname\relax
\expandafter\let\csname endequation*\endcsname\relax
\usepackage{amsmath}
\usepackage{float}
\usepackage{mathtools}
\usepackage{graphicx}
\usepackage{amsthm}
\usepackage{xcolor}
\usepackage[breaklinks=true,colorlinks=true,linkcolor=blue,urlcolor=blue,citecolor=blue]{hyperref}
\usepackage{soul}
\newcounter{enumeratecounter}

\newcommand{\half}{\frac{1}{2}}
\newcommand{\twothirds}{\frac{2}{3}}
\newcommand{\third}{\frac{1}{3}}
\newcommand{\threehalves}{\frac{3}{2}}
\newcommand{\sixth}{\frac{1}{6}}
\newcommand{\SL}{{\cal S}}
\newcommand{\F}{{\cal F}}
\newcommand{\C}{{\cal C}}

\newcommand{\omittext}[1]{{#1} } % or not

\DeclareMathOperator{\csch}{csch}

\begin{document}

\title[The cylinder amplitude in the Hard Dimer model on 2d CDT]{The cylinder amplitude in the Hard Dimer model on 2D Causal Dynamical Triangulations}

\omittext{
\author{John F Wheater}
\address{Rudolf Peierls Centre for Theoretical Physics\\
Department of Physics\\
Parks Road\\
Oxford OX1 3PU, UK}
\ead{john.wheater@physics.ox.ac.uk}

\author{P D Xavier}
\address{Rudolf Peierls Centre for Theoretical Physics\\
Department of Physics\\
Parks Road\\
Oxford OX1 3PU, UK}
\ead{praveen.xavier@physics.ox.ac.uk}}

\begin{abstract} We consider the model of hard dimers coupled to two-dimensional Causal Dynamical Triangulations (CDT) with all dimer types present and solve it exactly subject to a single restriction. Depending on the dimer weights there are, in addition to the usual gravity phase of CDT, two  tri-critical  and two dense dimer phases. We establish the properties of these phases, computing their cylinder and disk amplitudes, and their scaling limits.

\end{abstract}

\vspace{2pc}
\noindent{\it Keywords}: quantum gravity, low dimensional models, lattice models 
%\maketitle

\section{Introduction}

In this paper, we consider the hard dimer model on a fluctuating background in 1+1 dimensions. This background is provided by the ensemble of causal triangulations (CT) which triangulate  a surface $S^1\times R$ while maintaining a preferred time dimension. The  causal dynamical triangulation model (CDT) \cite{Ambjorn:1998xu} exists   in any dimension $D+1$ and was first suggested as an alternative to the model of dynamical triangulations (DT) \cite{DiFrancesco:1993cyw,Ambjorn:1997di} for discretized quantum gravity. DT works with unrestricted triangulations, or more generally planar random graphs, and naturally discretizes a space-time of Euclidean signature; the Lorentzian structure of the space-time in real quantum gravity does not  emerge in this picture. The CDT addresses this issue by using triangulations with a distinguished time direction (which can be Euclidean or Lorentzian). The causal structure of the discretized manifolds then survives in the continuum limit. A Hamiltonian operator and cylinder (or loop-loop) amplitude emerge   (see \cite{Zohren:2008vqi} for a review) which reproduce results first obtained in
\cite{Nakayama:1993np} by a proper-time gauge calculation in 1+1D continuum gravity. It has been shown \cite{Ambjorn:2013joa} that the continuum limit of CDT is equivalent to two-dimensional projectable Horava-Lifshitz gravity \cite{Horava:2009uw,Horava:2010zj}; we will refer to this as the `Pure gravity' (PG) phase of the extended models considered in this paper. For a review of the correspondence between CDT models and quantum gravity see \cite{Ambjorn:2012jv,Loll:2019rdj}.

Models in which extra degrees of freedom (often called `matter') are coupled to DT have been solved for many cases, showing that their scaling limit is equivalent to the corresponding flat lattice CFT coupled to Liouville gravity.  
Coupling matter to CDT is more  difficult and only limited progress has been made. Numerical work 
\cite{Ambjorn:2008jg,Ambjorn:2014pwa,Ambjorn:2014moa, Ambjorn:2015gea}  has established that the interaction between matter and geometry is weaker than for DT and the CDT with gauge fields is a solvable, but topological, system \cite{Ambjorn:2013rma}. The Ising model on CDT has been shown not to be magnetised at high enough temperature \cite{Napolitano:2015poa} but has otherwise resisted solution. It is known that the addition of curvature-squared \cite{DiFrancesco:1999em} or extrinsic curvature terms \cite{Glaser:2016smx} to the action does not change the universality class and recently loop models have been investigated \cite{Durhuus:2021tho}. 
However by exploiting the bijection of the CDT with a tree ensemble \cite{Durhuus:2009sm}, it is possible to make progress when the matter consists of hard dimers  \cite{Atkin:2012yt,Ambjorn:2012zx,Atkin:2012ka}. It was shown that a restricted form of the hard dimer (HD) model on CDT can be solved using a bijection with labelled trees; the model displays a transition away from the pure gravity phase of CDT, driven  by the dimer interactions. A slightly less restricted HD model was analysed in \cite{Ambjorn:2014voa} where a richer phase diagram was found. In this paper we show how to include all possible dimer types present on the CDT  (see Section \ref{sec:Hard Dimer Model}) in the bijection and  lift all the restrictions bar one which is required to render the tree system local. 

The tree bijection enables us to calculate the disk and cylinder amplitudes by reducing the problem to one of solving a non-linear second order recurrence relation. In contrast to the earlier work, with more restrictive dimer ensembles, these relations appear not always to be linearizable by a judicious mapping of the problem; they are inherently non-linear and cannot be solved analytically in closed form.   In the continuum limit however, the recurrence relations become a non-linear first or second order ODE, which takes a characteristic  form in each  phase of the model. The ODEs are vastly easier to solve than the recurrence relations and their solutions enable us to determine the continuum amplitudes and, in phases where it exists, a simple time translation operator.

This paper is organised as follows. We describe the CDT and HD models in sections \ref{sec:CDT Basics} and \ref{sec:Hard Dimer Model} respectively. Then we prove the bijection of a minimally restricted CDT+HD system with the labelled tree model in section \ref{sec:Bijection to Trees}. We construct the cylinder amplitude, and derive the defining recurrence relations in section \ref{sec:Cylinder Amplitude}. 
The fixed points of the recurrence relations determine the grand canonical partition function which we use to map out the phase diagram in section \ref{sec:Phases}. We then return to the cylinder amplitude and derive the basic ODEs that arise from the scaling limit of the recurrence relations, the continuum amplitudes and the associated time translation operators in sections \ref{scalinglimits} and \ref{Hamiltonian}. We conclude with a discussion of our results in Section \ref{sec:discussion}. 

\section{Causal triangulations}{\label{sec:CDT Basics}}

\begin{figure}[ht]
\centering
    \includegraphics[scale=0.4,trim={0 7.5cm 0 0},clip]{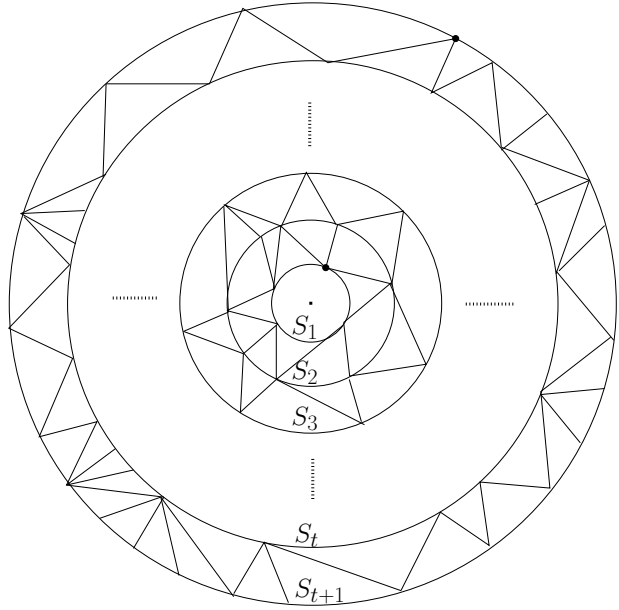}
    \caption{ \label{fig:CausalTriangulation} An element of $\mathcal{T}_t$. The marked vertices in $S_1$ and $S_{t+1}$ are highlighted in bold.}
\end{figure}

A causal triangulation $T  \in \mathcal{T}_t: t\in \mathbb{N}^{+}$ of the punctured disk and of height $t$ is defined as shown in Fig. \ref{fig:CausalTriangulation}. It consists of 

\begin{enumerate}
    \item A sequence of concentric circles $S_i$, $i\in[1,t+1]\cap \mathbb{N}$. 
    \item Edges connecting all vertices $v\in S_i$, $i\in[1,t]$ to at least one vertex $w\in S_{i+1}$  such that all faces are triangles.
    \item A marked vertex in $S_1$ and $S_{t+1}$.
\end{enumerate} 
We define the set of all causal triangulations of the punctured disk to be $\mathcal{T}:=\bigcup\limits_{t}\mathcal{T}_t$; the ensemble sum over 
$\mathcal{T}$ gives the CDT model. 

\begin{figure}[ht]
\centering
    \includegraphics[scale=0.4,trim={0 17cm 0 0},clip]{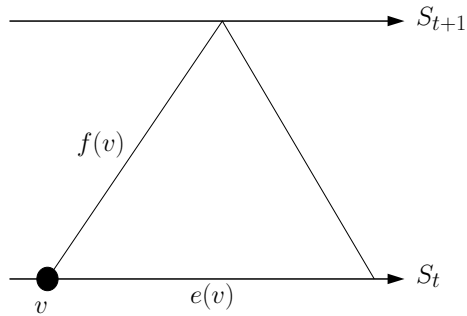}
    \caption{\label{fig:basics}Definitions of $f(v)$ and $e(v)$. The arrows point in the clockwise direction.}
\end{figure}

An edge connecting a vertex $v\in S_k$ to an immediately adjacent vertex $w\in S_k$ is said to be horizontal.
An edge is said to be forward (backward) directed with respect to a vertex $v \in S_k$ if it connects $v$ to a vertex in $S_{k+1}$ ($S_{k-1})$. A vertex $w\in S_k$ is said to be clockwise (anti-clockwise) -adjacent to a vertex $v\in S_k$, iff one meets $w$ first as one travels from $v$, clockwise (anti-clockwise) along $S_k$. 
For a given $T\in\mathcal{T}_t$, define (Fig. \ref{fig:basics}): 
\begin{enumerate}
    \item $e(v)$ as the horizontal edge connecting $v$ to the vertex that is clockwise-adjacent to $v$. The map $e$ is one to one and defined for all vertices in $T$.
    \item $f(v)$ as the forward directed edge w.r.t. $v$ which forms two of the edges of some unique triangle along with $e(v)$. The map $f$ is one to one and defined for all vertices in $\bigcup\limits_{i\in[1,t]}S_i$.
    \item $\sigma_{f}(v)$ to be the number of forward-directed edges w.r.t. $v$, defined for all vertices in $\bigcup\limits_{i\in[1,t]}S_i$.
    \item $\sigma_{b}(v)$ to be the number of backward-directed edges w.r.t. $v$, defined for all vertices in $\bigcup\limits_{i\in[2,t+1]}S_i$.
\end{enumerate}

\section{The restricted hard-dimer model} \label{sec:Hard Dimer Model}
\begin{figure}[ht]%[H]
    \centering
    \includegraphics[scale=0.4,trim={0 21.7cm 0 0},clip]{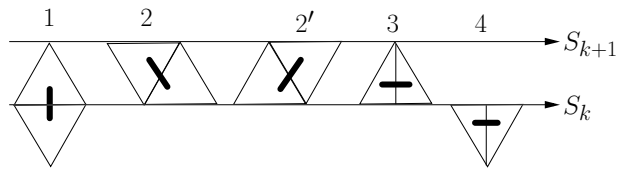}
    \caption{\label{fig:DimerTypes}The types of dimers. The clockwise direction is indicated by the arrow.}
\end{figure}

A hard dimer configuration on a triangulation is a configuration of short rods (aka dimers), each lying across an edge shared between two triangles (such an edge is said to be dual to a dimer), such that no two dimers are dual to edges of the same triangle and with the boundary condition that there are no dimers dual to boundary edges.
A list of all instances of occurrence of a dimer on a causal triangulation is shown in Fig. \ref{fig:DimerTypes}. In this paper we generalise and extend the work of \cite{Ambjorn:2014voa} by including dimers of type 4. We will solve the hard dimer model on causal triangulations subject to a single restriction which renders the model solvable and is a relaxation of that imposed in \cite{Ambjorn:2014voa}:

\noindent\textbf{Restriction 1:} If, for $v\in T$, $\sigma_b(v)>1$ and $\sigma_f(v)=1$, then $f(v)$ is not dual to a dimer. We call such a configuration `admissible'.

Let $\mathcal{D}_t: t\in\mathbb{N}^{+}$ denote the space of triangulations in $\mathcal{T}_t$ decorated by an admissible 
dimer configuration.
An element of $\mathcal{D}_t$ can be written as a pair $(T,D)$, where $T\in \mathcal{T}_t$ and $D$ is an admissible dimer configuration on $T$. Define $\mathcal{D}:=\bigcup\limits_{t\in\mathbb{N}^{+}}\mathcal{D}_t$.
The initial and final boundaries of a triangulation $T\in\mathcal{T}_t$ are given by
$\partial_1 T=S_1$ and $\partial_2 T=S_{t+1}$ respectively.

\begin{figure}[ht]
    \centering
    \includegraphics[scale=0.4,trim={0 22.7cm 0 0},clip]{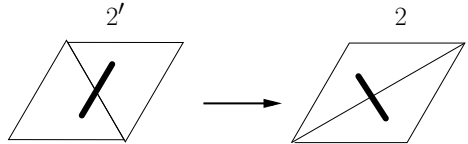}
    \caption{\label{fig:2primeto2Move}The 2' to 2 move.}
\end{figure}

\noindent The partition function (PF) of the set $\mathcal{D}_t$, often called the cylinder amplitude,  is defined by: 
\begin{equation}
G(g,x,y;\xi_{1},\xi_{2},\xi_{2'},\xi_{3},\xi_{4};t)=\sum_{(T,D)\in\mathcal{D}_t}x^{|\partial_1 T|}y^{|\partial_2 T|}g^{|T|}\xi_{1}^{|D|_1}\xi_{2}^{|D|_2}\xi_{2'}^{|D|_{2'}}\xi_{3}^{|D|_3}\xi_{4}^{|D|_{4}}, \label{eqn:1}
\end{equation}
where $|T|$, $|\partial_1 T|$, $|\partial_2 T|$ and $|D|_{i}$ count the number of triangles, edges on the initial and final boundaries, 
and dimers of type $i$ in $(T,D)$ respectively. 
Each dimer of type $i$ is weighted by a fugacity $\xi_i$, each triangle by a factor $g>0$,
each initial boundary edge by a factor $x>0$, and each final boundary edge by a factor $y>0$; if $g<g_c$, $x<x_c$, $y<y_c$, and the $\xi_i$ are not too negative, then the  series in
\eqref{eqn:1} is convergent.
As pointed out in \cite{Atkin:2012yt}, the move in Fig. \ref{fig:2primeto2Move} converts a dimer of type $2'$ to one of type $2$, without affecting the dimer configuration otherwise. Therefore, 
\begin{equation} G(g,x,y;\xi_{1},\xi_{2},\xi_{2'},\xi_{3},\xi_{4};t)=G(g,x,y;\xi_{1},(\xi_{2}+\xi_{2'}),0,\xi_{3},\xi_{4};t).
\end{equation}
So without loss of generality we set $\xi_{2'}=0$ and hereafter work with 
$G(g,x,y,\xi;t)$ where $\xi$ is the set of dimer weights $\{\xi_{1},\xi_{2},\xi_{3},\xi_{4}\}$. The amplitude for transition from an initial boundary of length $l_1$ to a final boundary of length $l_2$ in $t$ steps is given by the discrete inverse Laplace transform
\begin{equation}
    \tilde{G}(g,l_1,l_2,\xi;t)=\oint_\gamma\frac{du}{2\pi i u}\oint_\gamma\frac{dv}{2\pi iv}\, u^{l_1}\,v^{l_2}\,G(g,x_c/u,y_c/v,\xi;t), \label{eqn:Gtilde}
\end{equation}
where the contour $\gamma$ encloses the neighbourhood of the point at infinity in which  $G$ is analytic. 

\section{Bijection to labelled trees} {\label{sec:Bijection to Trees}}
    It was shown in \cite{Durhuus:2009sm} that there is a bijection between
    $\mathcal{T}_t$, the space of triangulations of the punctured disk with $t$ annuli, and the space of rooted plane branching trees of height $t+1$ denoted $\mathcal{P}_{t+1}$.
\begin{figure}[ht]
    \centering
    \includegraphics[scale=0.4,trim={0 21.3cm 8cm 0},clip]{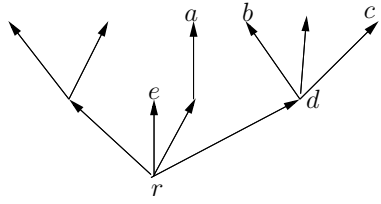}
    \caption{\label{fig:TreeTerminology}Tree terminology: the \emph{root} of the tree is $r$; the \emph{next} vertex of $b$ is $a$; the \emph{last child} of $d$ is $b$; the \emph{first child} of $d$ is $c$, and $e$ is a \emph{leaf} of the tree.}
\end{figure} 
We use the terminology for neighbouring sites on plane trees defined in Fig. \ref{fig:TreeTerminology}. 
Then the bijection $\beta$ maps $\mathcal{T}_t\ni T\to \tau\in\mathcal{P}_{t+1}$ as follows:
{\renewcommand{\labelenumi}{\Roman{enumi}}
\begin{enumerate}
\item Introduce a root vertex $r$ of $\tau$ in the centre of $S_1$ and add edges connecting it to each vertex in $v\in S_1\in T$. 
\item Delete $e(v)$ and $f(v)$ for all vertices $v\in T$.
\item The marked vertex in $S_1$ is the first child of the root vertex of $\tau$.
\end{enumerate}}
\noindent Aside from the root vertex, $\tau$ shares all its vertices with $T$. Bar the root vertex, we may therefore refer interchangeably to the vertices of $\tau$ and $T$. 

Now consider an element $(T,D)\in\mathcal{D}_t$. 
$T$ may be mapped to a plane tree $\tau=\beta(T)$ of height $t+1$. The dimer configuration $D$ induces a labelling $ \ell(v) $ of each vertex of $v\in\tau$ except the root where
\begin{equation}
    \ell(v)=(p(v),q(v)) \in \mathcal S =\{\{0,1,2,3\}\otimes \{0,4\}\setminus (1,4)  \}.
\end{equation}
We denote by $\mathcal{RL}_{t+1}=\gamma(\mathcal{D}_t)$ the set of labelled trees of height $t+1$ obtained from
$\mathcal{D}_t$ by applying $\beta$ and then assigning the labelling as follows:
\begin{enumerate}
\item \label{root} the root vertex is unlabelled;
\item \label{firstrule} if $\sigma_{b}(v)=1$ and this backward edge is dual to a dimer, 
$q(v)=4$; otherwise $q(v)=0$;
\item if $e(v)$ is dual to a dimer, $p(v)=1$;
\item if $f(v)$ is dual to a dimer and $\sigma_{f}(v)>1$, 
$p(v)=2$;
\item if $f(v)$ is dual to a dimer and $\sigma_{f}(v)=1$,
$p(v)=3$;
\item \label{lastrule} if neither $f(v)$ nor $e(v)$ is dual to a dimer,
$p(v)=0$.
\setcounter{enumeratecounter}{\value{enumi}}
\end{enumerate}
Not all labellings are allowed. They are constrained by:
\begin{enumerate}
\setcounter{enumi}{\value{enumeratecounter}}
    \item Boundary conditions, dimers dual to a boundary edge are not allowed: 
    \begin{enumerate}
    \item \label{itm:initial} if $v\in S_{1}$, then $p(v)\in \{0,2,3\}$ and $q(v)=0$;
        \item \label{itm:final} if $v\in S_{t+1}$, then $p(v)=0$.
    \end{enumerate}
    \item \label{harddimer} Hard dimer rule:
\begin{enumerate}
\item \label{itm:onefour} a vertex cannot have label $(1,4)$;
\item \label{itm:3pres} if a vertex $v$ has label $p(v)=3$, then the next (see Fig. \ref{fig:TreeTerminology} for the definition of \textit{next}) vertex $w$ has $p(w)=0$;
\item \label{itm:2des} if a vertex $v$ has label $p(v)=2$, then the first child $w$ cannot have $p(w)=1$ or $q(w)=4$;
\item \label{itm:4pres} if a vertex $v$ has label $q(v)=4$, then the next vertex $w$ cannot have $p(w)=1$ or $q(w)=4$.

\end{enumerate}
        \item \label{itm:3left} Restriction: if vertex $v$ is a last child, it cannot have $p(v)=3$.
        \item \label{geometry}Geometry:
        \begin{enumerate}
\item \label{itm:leaf} a vertex $v$ with label $p(v)=3$ is necessarily a leaf;
\item \label{itm:2atleastone} a vertex $v$ with label $p(v)=2$ has at least one child; 
\item \label{itm:4leftmost} a vertex $v$ with label $q(v)=4$ is not a last child.
\end{enumerate}
\end{enumerate}
The map $\gamma$ obtained by composing $\beta$ with the rules
(\ref{root}) through (\ref{geometry}) is a bijection from $\mathcal{D}_t$ to $\mathcal{RL}_{t+1}$. The proof is given in \ref{bijectionproof}. We define $\mathcal{L}^n_t$ to be the set of all labelled trees of height $t$ with a root label $n=\ell(r)\in\mathcal{S}$ and  satisfying all the above constraints \emph{except} the initial boundary condition vii(a); and finally the set
$\mathcal{L}^{n}_{\leq t}=\bigcup\limits_{i\in[0,t]}\mathcal{L}^{n}_i$.

All conditions on the labelling, except the restriction (\ref{itm:3left}), are consequences of the full hard dimer model; they arise either from the hard dimer rule, or from the geometry of $T\in{\cal T}_t$ and its associated tree. 
The restriction ensures that interactions in the labelled tree model are local; otherwise
if $v$ is a last child with $p(v)=3$, then, by the hard dimer rule \ref{itm:3pres}, the label of the next vertex, which is not a sibling of $v$, is constrained thus generating a non-local interaction in the tree.
Note that the model with no dimers of type 3, equivalently $\xi_3=0$, is not changed at all by the restriction.

Now consider the inversion operation $I: \mathcal{T}_t\to \mathcal{T}_t$ defined as $S_k(I(T)):=\mathcal{R}(S_{2+t-k}(T))$ with the connectivity of vertices unchanged, and where the $\mathcal{R}$ operation $\mathcal{R}(S_k)$ reverses the orientation of the circle $S_k$. Under $I$, the dimer types are mapped $1\to 1$, $2\to 2$ and $3\leftrightarrow 4$, while the boundary fugacities are swapped $x\leftrightarrow y$. 
Under $I$, the restriction \ref{itm:3left} on $T$, becomes the restriction that a vertex with label $q=4$ in $I(T)$, is necessarily a leaf. Conversely, a vertex in $T$ with label $q=4$ that is \underline{not} a leaf gets mapped to a vertex that is a last child assuming the label $p=3$ in $I(T)$. It follows that while the full dimer model is invariant under $I$, this symmetry is broken by the restriction; this fact will be important later. 
A further corollary is that the 
unrestricted model with $\xi_4=0$   can in fact be  solved by the methods of this paper, since it maps under $I$ to our model with $\xi_3=0$, for which the restriction is empty. 

\section{Generating Functions}
\begin{figure}[ht]
    \centering
    \includegraphics[scale=0.4,trim={0 17.5cm 0cm 0},clip]{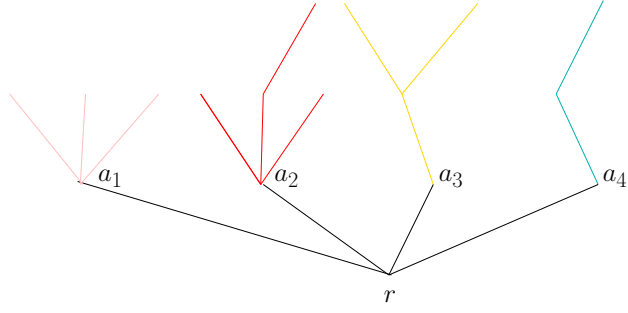}
    \caption{Decomposing a tree into sub-trees of the root vertex. $a_{1,2,3,4}$ are the roots of the sub-trees of $r$. \label{fig:Subtrees}}
\end{figure}

{\label{sec:Cylinder Amplitude}}
By mapping $\mathcal{D}_t$ to $\mathcal{RL}_{t+1}$ with the bijection $\gamma$, we find that
\begin{equation}
    G(g,x,y,\xi;t)=\sum_{(\tau,l)\in\mathcal{RL}_{t+1}}g^{2|\tau|-2}(x/g)^{|\partial_1 \tau|}(y/g)^{|\partial_2 \tau|}\prod_{v\in\tau\setminus r}\xi_{p(v)}\xi_{q(v)}, %
    \end{equation}
where $|\tau|$ is the number of vertices of $\tau$, $|\partial_1 \tau|$ is the number of vertices at height $1$, $|\partial_2 \tau|$ is the number of vertices at height $t+1$, and for convenience we define $\xi_0\equiv 1$.
We define the PF of $\mathcal{L}^{n}_{\leq t}$ to be
\begin{align}
    &W^n(g,y,\xi;t)=\sum_{(\tau,l)\in\mathcal{L}^{n}_{\leq t}}g^{2|\tau|}(y/g)^{|\partial_2 \tau|}
    \prod_{v\in\tau}\xi_{p(v)}\xi_{q(v)}.
    \end{align}
where $|\partial_2 \tau|$ here means the number of vertices at height $t$ and $|\tau|$ remains as defined above. By the dimer rule vii(a), only labels $(0,0), (2,0), (3,0)$ are for allowed for vertices on $S_1$. Decomposing a tree in $\mathcal{RL}_{t+1}$ in terms of sub-trees in $\mathcal{L}^{n}_{\leq t}$  (see Fig. \ref{fig:Subtrees}) \cite{Atkin:2012yt}\cite{Ambjorn:2014voa} we find that
\begin{multline}
    G(g,x,y,\xi;t)= \\ y\partial_y\, 
    \frac{1}{1-\frac{x}{g}(W^{(0,0)}(g,y,\xi;t)(1 +\frac{x}{g}W^{(3,0)}(g,y,\xi;t))+W^{(2,0)}(g,y,\xi;t))},
        \label{eqn:h0}
\end{multline}
thus relating the partition function of $\mathcal{RL}_{t+1}$ to those of $\mathcal{L}^{n}_{\leq t}$.  
    The operation $y\partial_y$ marks the exit loop and removes the set of trees which are of height less than $t$. The model can be solved by finding recursion relations for the $W^n$ to which we now turn.
    
    The initial conditions are given by trees of height zero. By the dimer rule vii(b), only labels $(0,0)$ and $(0,4)$ are allowed on the final boundary so
    \begin{align}
        W^{(0,q)}(g,y,\xi;0) &=yg\xi_q, \nonumber\\
        W^{(p\ne 0,q)}(g,y,\xi;0)&=0.\label{eqn:initial}
    \end{align}
A vertex with label $(3,0)$ is necessarily a leaf (see \ref{itm:leaf}). Therefore,
\begin{equation}
    W^{(3,0)}(g,y,\xi;t+1)= g^2 \xi_{3},\qquad t\geq 0. \label{eqn:h4}
\end{equation}
The $W^n$ are not all independent.
Since there are no restrictions on the children of a vertex with label $(1,0)$,
\begin{align}
W^{(1,0)}(g,y,\xi;t+1) 
    = \xi_1 W^{(0,0)}(g,y,\xi;t+1), \qquad t\geq 0.\label{eqn:h2}
\end{align}
Finally, since the label $q(v)=4$ does not place any restrictions on the children of the vertex $v$, for $t\geq0$ we have
\begin{equation}
    W^{(p,4)}(g,y,\xi;t+1)
    = \xi_4 W^{(p,0)}(g,y,\xi;t+1), \qquad t\geq 0.\label{eqn:h567}
\end{equation}
Trees of height of at least 1 can be decomposed 
into the sub-trees of the root vertex
(see Fig. \ref{fig:Subtrees}). Applying the dimer labelling rules we find the recursions
\begin{multline}
   W^{(0,0)}(g,y,\xi;t+1)=  \\
   \frac{ g^2}{1-\left( W^{(1,0)}+W^{(0,0)}W^{(3,4)}
    +\frac{\left(W^{(0,0)}W^{(3,0)}+W^{(0,0)}+W^{(2,0)}\right)\left(1+W^{(0,4)}+W^{(2,4)}\right)}{1-W^{(3,0)}W^{(0,4)}}\right)}, \label{eqn:h1}
\end{multline}
and
\begin{multline}
    W^{(2,0)}(g,y,\xi;t+1)=  \\
    \frac{ g^2\xi_2\left(\frac{W^{(0,0)}+W^{(  2,0)}+W^{(0,0)}W^{(3,0)}}{1-W^{(0,4)}W^{(3,0)}}\right)}{1-\left( W^{(1,0)}+W^{(0,0)}W^{(3,4)}
    +\frac{\left(W^{(0,0)}W^{(3,0)}+W^{(0,0)}+W^{(2,0)}\right)\left(1+W^{(0,4)}+W^{(2,4)}\right)}{1-W^{(3,0)}W^{(0,4)}}\right)},  \label{eqn:h3}
\end{multline}
for $t\geq 0$. Here  all $W^n$ on the right-hand-side are evaluated at
\noindent $\left(g,y,\xi;t\right)$. Recursions for all other $W^n$ follow from these two by using the relationships 
\eqref{eqn:h4}, \eqref{eqn:h567}.

It is convenient to define:
\begin{align}
    f^1_t(y)&=\frac{1}{g^2} W^{(0,0)}\left(g,y,\xi;t\right),\qquad  t\geq 1,\nonumber\\
    f^2_t(y)&=\frac{1}{\xi_2 g^2} W^{(2,0)}\left(g,y,\xi;t\right),\qquad  t\geq 1.
\end{align}   
Then, applying \eqref{eqn:h4}-\eqref{eqn:h567}, 
to (\ref{eqn:h1}) and (\ref{eqn:h3}), we find that  
the $f^i_t(y)$ satisfy the initial conditions
\begin{align}
    f^1_1&=\frac{1}{1-gy-g^2y^2\xi_4}, \label{f0n} \\
    f^2_1&=\frac{gy}{1-gy-g^2y^2\xi_4}, 
    \label{eqn:initialconditions}
    \end{align}
and the recursion relations   (the $y$ argument is suppressed for brevity),
    \begin{align}
    f^1_t&=\F^1(f^1_{t-1},f^2_{t-1},g)\equiv\frac{1}
    {1-A_{t-1}}, \label{eqn:scale1}\\
    f^2_t&=\F^2(f^1_{t-1},f^2_{t-1},g)\equiv\frac{1}{1-A_{t-1}}\left(\frac{g^2f^1_{t-1}+g^2\xi_2 f^2_{t-1}+g^4\xi_3f^1_{t-1}}{1-g^4\xi_3\xi_4f^1_{t-1}}\right), \label{eqn:scale2} \\
    A_{t}&=g^2\xi_1f^1_t+g^4\xi_4\xi_3f^1_t+\frac{g^2(g^2\xi_3f^1_t+f^1_t+\xi_2f^2_t)(1+g^2\xi_4f^1_t+g^2\xi_4\xi_2f^2_t)}{1-g^4\xi_3\xi_4f^1_t}. \label{thend}
\end{align}
Finally, from (\ref{eqn:h0}) we find that
\begin{equation}
   G(g,x,y,\xi;t)=y\,\partial_y\frac{1+g^2x^2\xi_3f^1_{t}}{1-g\,x(\xi_2\, f^2_{t}(y)+(1+g\,x\,\xi_3)f^1_{t}(y))}.\label{eqn:amp}
\end{equation}

\section{Phases and Critical Exponents}{\label{sec:Phases}}

The Grand Canonical partition function is determined by $f^1_\infty$ which is the fixed point of the sequence generated by \eqref{f0n}-\eqref{thend} i.e. $f^1_{t-1}=f^1_t=f^1_{t+1}=f^1_{\infty}$.  Eliminating $f^2_{\infty}$ from the coupled equations 
\eqref{f0n}-\eqref{thend} we find that 
$P(f^1_\infty)=0$ where $P(.)$ is a quartic.
Provided the dimer fugacities lie in the physical region above a limit surface ${\SL}(\xi)=0$, on which at least one dimer fugacity must be negative, the dimer dynamics is not critical and does not significantly affect the geometry which remains that of pure gravity. For fugacities below the limit surface the model does not exist in the statistical mechanical sense. 
On the limit surface itself the dimers can be  critical and, as in \cite{Ambjorn:2014voa}, can then interact with the geometry strongly enough to drive it away from the pure gravity behaviour. 
In the rest of this section we discuss these properties in more detail.

\subsection{Dimer phases}

For a fixed set of fugacity values $\xi$ in the physical region, the expansion (\ref{eqn:1}) converges and $f^i_\infty(g,\xi)$ are analytic functions for $0<g<g_c(\xi)$.   At $g=g_c(\xi)$, the physical solution  $f^i_\infty(g_c,\xi)=f^i_c(\xi)$ is either a double root of $P$, or a triple root of $P$ 
in which case the system is `tri-critical' (a physical quadruple root never occurs).  Close to $g=g_c(\xi)$ we expect $f^i_\infty(g,\xi)$ to
exhibit the scaling behaviour
\begin{equation}
    \Delta f^i=f^i_\infty(g,\xi)- f^i_c(\xi)= - \phi^i_c(\xi)(\Delta g)^\alpha + h.o.t. \label{fscaling}
\end{equation}
 where $\Delta g= g_c(\xi)-g$. In the bulk of the physical region, which contains the pure gravity model $\xi_i=0, \forall i$,  there is a double root at $g=g_c(\xi)$; the dimers do not interact strongly with the geometry and $\alpha=\half$.  On the boundary of the physical region there is either a double root, in which case $\alpha=\half$, or the dimers interact strongly with the geometry driving the system tri-critical and generically $\alpha =\third $. 
 The free energy density of the system is given by 
\begin{equation}
    \mu(\xi)=-\log g_c(\xi).
\end{equation}
 In the bulk of the physical region
 $g_c(\xi)$ is a single real root 
 of the discriminant of $P(f)$ with respect to $f$
 so the dimer density is an analytic function of the fugacities. However $g_c(\xi)$ is a multiple root on the surface $\xi_{CD}$ where it is a non-analytic function of $\xi$
 \begin{equation}
      g_c(\xi)=g_c(\xi_B) + R_1(\xi)+(\Delta\xi)^{1+\sigma} g^{(1)}_c(\{\xi_{}\}_B)(1+R_2(\{\xi\})).
 \end{equation}
 Here $\Delta\xi$ is defined by $\{\xi_i=\xi_{iCD}+\Delta \xi, \forall i\}$,  and $R_{1,2}$ are a regular functions vanishing on $\xi_{CD}$.  In this model $\sigma$ takes value  $-\third$, in which case the dimer density 
 \begin{equation}
     \rho_{\rm dimer}=\xi\,\partial_\xi \mu\sim (\Delta\xi)^{\sigma}.
 \end{equation} 
 diverges, or $\half$.

 By considering the flow of roots we can see that $\xi_{CD}$ forms part of the limit surface $\SL$. First consider the double root case. As a given fugacity $\xi_i$ is decreased to $\xi_{iCD}$ while $\xi_{j\ne i}$ are held fixed at their $\xi_{CD}$ values
 the physical root $g_c(\xi)$ meets another real root; then as $\xi_i$ is decreased further the two roots flow out into the complex plane and the statistical mechanical interpretation is lost.  A similar argument applies for the triple root case. 
  
\subsection{Phase conditions}
To establish in detail the behaviour of $f^i_\infty(g,\xi)$ close to the critical surface, it is more convenient to study the relations \eqref{eqn:scale1}, \eqref{eqn:scale2} directly.
Differentiating  we find that (from now on, we will usually suppress the ${\xi}$ arguments)
\begin{equation}  (1-\mathbb{T}^{ij})\frac{\partial f^j_\infty}{\partial g}=\frac{\partial \F^i}{\partial g},
\end{equation}
where 
\begin{equation}  \mathbb{T}^{ij}  =\frac{\partial \F^i}{\partial f^j}\equiv \F^i_j\label{Tmatrix}
\end{equation}
plays a special role in what follows.
At the critical surface, ${\mathbb{T}}$ has (at least) one eigenvalue $\lambda_1=\lambda_{1c}=1$ and furthermore, by definition, no eigenvalue can reach 1 for $g<g_c(\xi)$. 
Defining
\begin{equation} \C(\lambda)=\rm{det} (\lambda\mathbb{I}- \mathbb{T})=\F^1_1 \F^2_2 - \F^1_2\F^2_1 -\lambda(\F^1_1+\F^2_2) +\lambda^2,
\end{equation}
the criticality condition is that $\C(1)=0$ at $g=g_c(\xi)$,
and the second eigenvalue is given by
\begin{equation} \lambda_{2c}= \F^1_1+\F^2_2 -1.
\end{equation}
Provided $\F^1_1\ne 1,\,\F^1_2\ne 0$, the corresponding eigenvectors are
\begin{equation}
   u_{1c}= \begin{pmatrix}\F^1_2\\1-\F^1_1\end{pmatrix},\qquad u_{2c}= \begin{pmatrix}\F^1_2\\  \F^2_2-1\end{pmatrix},\label{eqn:eigenvectors}
  \end{equation} 
  and  vectors orthogonal to $u_{1c}$ and $u_{2c}$, which we denote 
  $\bar u_{1c}$ and $\bar u_{2c}$ respectively, are
   \begin{equation}
   \bar u_{1c}= \begin{pmatrix}1-\F^1_1 \\-\F^1_2\end{pmatrix},\qquad \bar u_{2c}= \begin{pmatrix}\F^2_2-1\\-\F^1_2\end{pmatrix}.\label{eqn:eigenvectorduals}
  \end{equation} 
  Note that if $\lambda_{2c}=1$ then  $u_{2c}=u_{1c}$. In this case $\mathbb{T}$ is not diagonalizable but takes a Jordan normal form at $g=g_c(\xi)$; there is one regular eigenvector $u_{1c}$ with eigenvalue $\lambda_1=1$, and a linearly independent vector $u_{2c}$ satisfying
\begin{equation}
    {\mathbb{T}}u_{2c}=u_{2c}+\epsilon\, u_{1c},
\end{equation}
where $\epsilon$ is a function of the dimer fugacities $\xi$.

Now we expand \eqref{eqn:scale1} and \eqref{eqn:scale2} about the critical point  by setting
\begin{eqnarray}
    f^i_{\infty}&=&f^i_{c}+\phi^i,\nonumber\\
    g&=&g_c-\Delta g,
\end{eqnarray}
 which gives  
\begin{eqnarray} (1-{\mathbb{T}})^{ij}\phi^j&=& -\Delta g\left(\frac{\partial{\F^i}}{\partial g}+\frac{\partial {\mathbb{T}}^{ij}}{\partial g}\phi^j\right)
+\frac{1}{2}\F^i_{\ell k}\, \phi^\ell\phi^k\nonumber\\ &&+\frac{1}{3!}\F^i_{k\ell m}\phi^k\phi^\ell\phi^m +O\left((\Delta g)^2,\phi^4\right),\label{eqn:master}
\end{eqnarray}
and then decompose $\phi^i$ in the $u_{1c}, u_{2c} $ basis.  Various phases occur as some coefficients in \eqref{eqn:master} vanish which in turn causes different behaviour for $\phi^i$. There are three cases (here $\phi_c>0$ and $\chi_c$ are constants that depend on the dimer fugacities):
\begin{enumerate}
    \item Generic 
\begin{equation}
    \bar u^i_{2c}\frac{\partial{\F^i}}{\partial g}\ne 0,\quad  \bar u^i_{2c}(u^\ell_{1c} \F^i_{\ell k}u^k_{1c}) \ne 0, 
\end{equation}
which leads to
\begin{equation}
    \phi^i=-\phi_c (\Delta g)^\frac{1}{2}  u^i_{1c} -\chi_c(\Delta g)\,u^i_{2c}+h.o.t.\label{eqn:phinfg}
\end{equation}
\item Tri-critical 
\begin{equation}
    \bar u^i_{2c}\frac{\partial{\F^i}}{\partial g}\ne 0,\quad  \bar u^i_{2c}(u^\ell_{1c} \F^i_{\ell k}u^k_{1c}) = 0, 
    \label{tricritcond}
\end{equation}
which leads to 
\begin{equation}
    \phi^i=-\phi_c (\Delta g)^\frac{1}{3}  u^i_{1c} -\chi_c (\Delta g)^\frac{2}{3}\, u^i_{2c}+h.o.t.\label{eqn:phinftc}
\end{equation}
\item Dense Dimer (so-called because it turns out that this phase has $\sigma<0$)% is
\begin{equation}
    \bar u^i_{2c}\frac{\partial{\F^i}}{\partial g}= 0,\quad  \bar u^i_{2c}(u^\ell_{1c} \F^i_{\ell k}u^k_{1c}) = 0, \label{ddconstraint}
\end{equation}
which leads to 
\begin{equation}
    \phi^i=-\phi_c (\Delta g)^\frac{1}{2}  u^i_{1c} -\chi_c(\Delta g)\, u^i_{2c}+h.o.t.\label{eqn:phinfdd}
\end{equation}
\end{enumerate}

\subsection{The Hausdorff dimensions}

The exponents $\alpha$ and $\sigma$ are not alone sufficient to characterise the phase diagram -- we also need measures of the geometry. 
The global and local Hausdorff dimensions, $d_H$ and $d_h$, are the simplest characterisation of the geometrical properties of these discretized systems. They measure 
the growth with height $t$ of the spatial size of the triangulations.  

The global Hausdorf dimension is determined through the large $t$ behaviour of the two point function defined by
\begin{eqnarray}
    {\mathbb G}_{ij}(g,{\xi};t)&=&
    \sum_{t'=t}^\infty\sum_{(\tau,l)\in\mathcal{L}^i_{t'+1}: \ell(v_1)=j}   g^{2|\tau|}
    \xi_1^{l_1}\xi_2^{l_2}\xi_3^{l_3}\xi_4^{l_4},\label{def:G}
    \end{eqnarray}
where $v_1$ is a marked vertex at height $t$ (see \cite{Ambjorn:1997di} for details). It is sufficient to consider the  two dimensional sub-matrix  with $  \ell,\ell'\in \{(0,0),(2,0)\} $ so, noting that the sum runs over all trees of height at least $t+1$,  
\begin{equation}
      {\mathbb G}_{ij}(g,{\xi};t)=({\mathbb T}^t)_{ij}.
\end{equation}
As $t\to\infty$ this takes the form, as $g\uparrow g_c$,
\begin{equation}
    {\mathbb G}_{ij}(g,{\xi};t) = G_{ij}\, e^{-m(g)t +o(t)},
\end{equation}
where $G_{ij}$ is a constant, and the global Hausdorff dimension $d_H$ is defined by
\begin{equation}
    m(g)=c\, (\Delta g)^{\frac{1}{d_H}}+ h.o.t.
\end{equation}
On the other hand the large $t$ behaviour of $\mathbb G_{ij}$ is determined by the eigenvalues $\lambda_{1,2}$ that take the value 1 at $g=g_c$. 
Defining the correlation exponent $\nu$ through
\begin{equation}
   \lambda=1- \Delta \lambda = 1- b (\Delta g)^\nu + h.o.t.,
\end{equation}
where $b>0$ is a constant, we see that $d_H= \nu^{-1}$ \cite{Ambjorn:1997di}. The correlation function 
has a scaling limit for large times; letting $t=Ta^{-1}$, $\Delta g = \Lambda a^\frac{1}{\nu} = \Lambda a^{d_H}$ and taking $a\to 0$ we have
\begin{equation}
    {\mathbb G}^s_{ij}(\Lambda,{\xi};T)=\lim_{a\to 0}  {\mathbb G}_{ij}(g_c-\Lambda a^{d_H},{\xi} ;T/a) =G_{ij}\, e^{-b\Lambda^\nu T}.\label{eqn:TwoPtFnScaling}
\end{equation}
For $g<g_c$ the eigenvalue $\lambda_1=1-\Delta \lambda_1$ where $\Delta\lambda_1$ satisfies
\begin{equation}
    0=\Delta \lambda^2+\Delta \lambda\, (\lambda_{2c}-1+\F^i_{ij}\Delta f^j) +\partial_i \C(1)\, \Delta f^i+ \half \partial_i\partial_j \C(1) \Delta f^i \Delta f^j+O(\Delta g,(\Delta f) ^3).\label{Lambdascaling}
\end{equation}
On the critical surface (ie when $\C(1)=0$) it is straightforward to show that
\begin{equation}
\partial_j \C(1)=(\F^1_2)^{-1}  \bar u^i_{2c}(u^\ell_{1c} \F^i_{\ell j}),
\end{equation}
so if the tri-critical condition \eqref{tricritcond} is satisfied, the term linear in $\Delta f^j$ is
actually the same order in $\Delta g$ as the quadratic term. There are then four possible cases for $\Delta\lambda$:
\begin{enumerate}
 \item Generic: the system is not tri-critical, and $\lambda_{2c}<1$. In this case the dimers do not interact strongly enough with the geometry to affect its large scale properties and 
     \begin{equation}
        \Delta\lambda=b (\Delta g)^\frac{1}{2} +h.o.t.
    \end{equation}
    \item \label{degenerate} Degenerate: the system is not tri-critical, \emph{but} $\lambda_{2c}=1$. We find
     \begin{equation}
        \Delta\lambda_{1,2}=\pm i\, b (\Delta g)^{\half\alpha} +h.o.t.
            \end{equation}
   
    \item Tri-critical I: the system is tri-critical, \emph{and} $\lambda_{2c}=1$. 
    We find 
    \begin{equation}
        \Delta\lambda_{1,2}=\omega_{1,2}(\Delta g)^\third + h.o.t.
    \end{equation}
    where 
    $\omega_{1,2}$ satisfy
    \begin{equation}
        0=\omega^2-\omega (\F^i_{ij}u^j_{1c}\phi_c)  + \half \partial_i\partial_j \C(1)\,\phi_c^2 u^i_{1c}u^j_{1c}-\partial_i \C(1) \chi_c u^i_{2c},\label{eqn:omega}
    \end{equation}
    and direct calculation shows that \begin{equation}\omega_{1,2}=\frac{3}{2f^1_{c}}(1\pm i/\sqrt{3}).\label{omegaval}\end{equation}

    \item Tri-critical II:  the system is  again tri-critical, but $\lambda_{2c}<1$ which gives
        \begin{equation}
        \Delta\lambda=b\,(\Delta g)^{\frac{2}{3}} + h.o.t. 
    \end{equation}

\end{enumerate}

The local Hausdorff dimension $d_h$ is defined by first setting $g=g_c$. The tree ensemble generated at $g=g_c$ then has  a probabilistic interpretation provided the dimer fugacities are not too negative \cite{Ambjorn:2014voa}. The limiting ensemble as system size $N\to\infty$ consists of infinite trees $\{\tau\}$ with a single infinite spine and a probability measure $\mu(\tau)$.  The expectation value of the volume (equivalently the number of vertices) 
the  ball $B_R$ of radius $R$ centred on the root then determines $d_h$ through
\begin{equation}
   \langle \vert B_R\vert \rangle_\mu= c\, R^{d_h}(1+O(R^{-1})).
\end{equation}
In this case the value of $d_h$ is determined by the structure  of $\mathbb T$ and  the second derivatives $\F^i_{jk}$
at $g=g_c$ and can be computed using the methods of the appendix of \cite{Ambjorn:2014voa}.

\subsection{The Phase Diagram}\label{subsec:phasediag}

\begin{table}[ht]
   \centering
    \begin{tabular}{l|c|c|c|c|c|c}
    Dimer state& $\lambda_{2c}$ &$\alpha$&$\sigma$&$\nu$&$d_H$&$d_h$\\\hline
       Tri-critical I& $1$  & $\third$&$\half$ &$\third$ & 3 &3 \\
        Dense-dimer I& $1$  & $\half$&$-\third$ & $\half$& 2 &3 \\
         Tri-critical II& $<1$ & $\third$ &$\half$ & $\twothirds$& $\threehalves$& 1\\
        Dense-dimer II& $<1$  & $\half$&$-\third$ & $1$& 1 &1 \\
                Pure-gravity& $<1$&$\half$&n/a &$\half$ &2&2
    \end{tabular}
    \caption{Exponents and Hausdorff dimensions for the possible phases of the model. }
    \label{tab:Table1}
\end{table}

\begin{figure}[ht]
    \centering
    \includegraphics[scale=0.6,trim={0 0cm 0 0},clip]{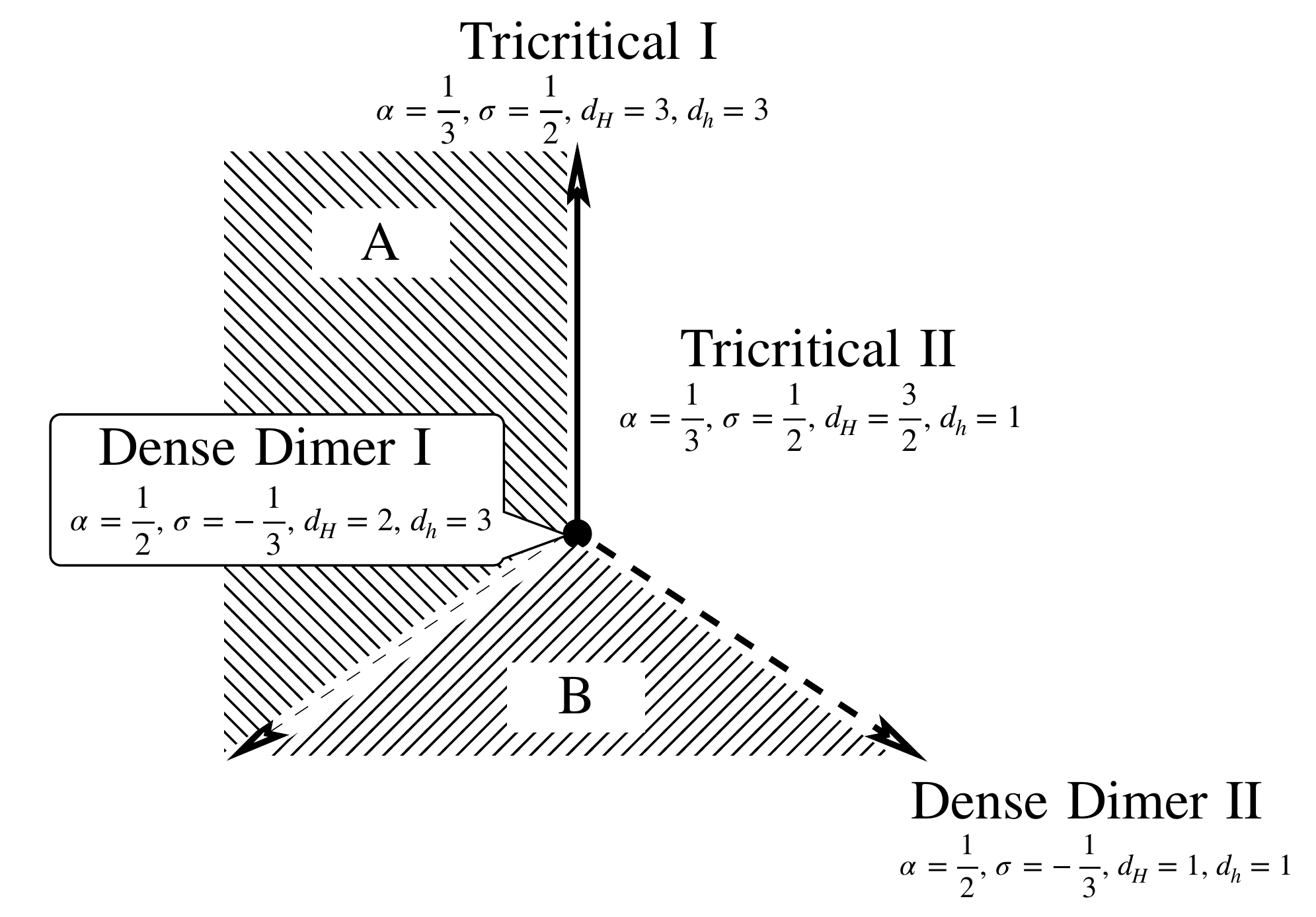}
    \caption{\label{fig:NewPhaseDiagram} The limit surface $\SL$, restricted to $\xi_1=\xi_2=\xi$, showing all the non pure-gravitational phases of the model. Note that the shaded regions  are not distinct phases.}
\end{figure} 
The full form of the limit surface $\SL$ and the phase diagram for general $\xi$ is complicated and not very informative. The regimes that occur in the model and the corresponding values of exponents and Hausdorff dimensions are listed in Table \ref{tab:Table1}. Above the limit surface $\SL$ the dimers are not critical and the system is in the regular Pure-gravity phase.  Restricting to $\xi_1=\xi_2=\xi$, the limit surface is two dimensional and is shown schematically in Fig \ref{fig:NewPhaseDiagram}. The relative richness of the phase diagram is a consequence of the structure of the CDT graphs enabling us to keep track of different orientations of dimers and assign different fugacities to them -- something which cannot be done in the planar random graph case for example. 
We note the following:
\renewcommand{\labelenumi}{\alph{enumi})}
\begin{enumerate}
  \item The Tri-critical I (TCI) line lies in the plane $\xi_4=0$.    The degenerate  eigenvalues of $\mathbb T $ lead to the oscillatory behaviour superimposed on exponential decay of $\mathbb G_{ij}(g,\xi;t)$  \eqref{omegaval}, and to $d_H=3$ which violates the naive scaling prediction $d_H=\threehalves$. 
   This  phase was first identified in \cite{Atkin:2012yt,Ambjorn:2012zx} and the unexpected Hausdorff dimension in \cite{Ambjorn:2014voa}. 
     \item The Tri-critical II (TCII) phase occurs when $\xi_4<0$,
     and represents the generic tri-critical behaviour of the model with Hausdorff dimension $d_H=\threehalves$ in agreement with hyperscaling. It contains the point with all dimer fugacities equal, $\xi_i\simeq -0.1646$, and also the point $\xi_{1,2,3}=0,\,\xi_4=-\frac{1}{3}$
     for which the restriction (\ref{itm:3left}) has no effect; we will exploit this in the next section.
          \item The Dense-dimer I (DDI) point terminates the TCI line (at  $\xi\simeq-0.2267,\,\xi_3\simeq-0.2781$, $\xi_4=0$), and       the Dense-dimer II (DDII) line which emerges in the $\xi_4<0$ direction.%
     
      \item   The dimer density exponent for the TC phases, $\sigma=1/2$,  
    is the same as that found in the model of dimers coupled to planar random graphs \cite{Staudacher:1989fy}, and in earlier studies of dimers coupled to CDTs or trees \cite{Atkin:2012yt,Ambjorn:2012zx,Ambjorn:2014voa}. 
      The planar random graph result was a puzzle because it appears to contradict the KPZ formula \cite{Knizhnik:1988ak}; inserting the value for the regular square lattice dimer model, $\sigma=-\frac{1}{6}$ \cite{Cardy:1985yy}, into the formula gives $\sigma=-\third$. This phenomenon has been explained in terms of operator mixing between geometric and matter degrees of freedom by \cite{Ambjorn:2014jca}. 
      On the other hand, the DD phases do have $\sigma=-\third$. It appears that with extra dimer fugacities available we can fine-tune so that the coefficient in the mixing vanishes thus revealing the $\sigma=-\third$ behaviour. 
    \item In region A, both eigenvalues $\lambda_{1,2} =1 $ at $g=g_c$ but $\Delta \lambda$ is imaginary so     the leading behaviour of ${\mathbb G}_{ij}$ is pure oscillatory and $d_H$ does not exist. We can understand why the surface, $\xi_{D}$, on which this occurs  forms part of $\SL$ by considering the flow of $\lambda_{1,2}$
as $g\uparrow g_c$. Holding dimer fugacities $\xi_{j\ne i}$ fixed at their $\xi_{CD}$ values while varying 
 $\xi_i$ through $\xi_{iCD}$ gives the flows shown in Fig. \ref{fig:lambda}. For $\xi_i<\xi_{iCD}$,  $\vert \lambda_{1,2}\vert$ exceeds 1 when $g<g_c$, resulting in exponential growth of correlation functions. As observed in \cite{Ambjorn:2014voa} this phenomenon indicates failure of absolute convergence of the series (\ref{def:G}).  \begin{figure}[ht]
      \centering
    \includegraphics[scale=0.4,trim={0 0cm 0 0},clip]{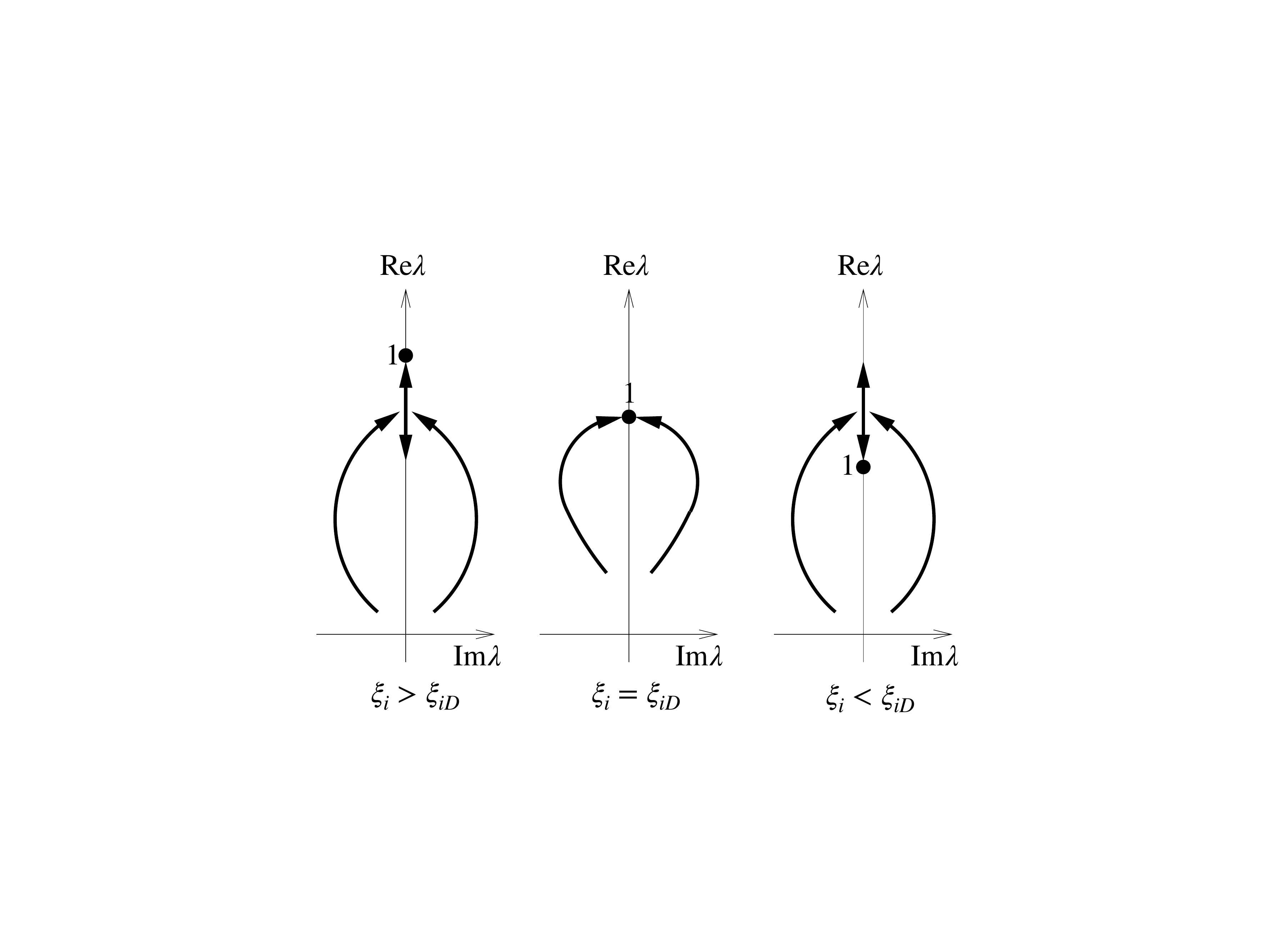}
    \caption{\label{fig:lambda} Flow of $\lambda_{1,2}$
as $g\uparrow g_c$. The arrows denote the direction of flow as $g$ increases. }
\end{figure}

    \item In region B small graphs dominate so there is no thermodynamic limit and the statistical mechanical interpretation fails. 
   
\end{enumerate}

\section{Scaling limit of the Cylinder Amplitude \label{scalinglimits}}
\subsection{Introduction}

We now return to the micro-canonical ensemble of surfaces with fixed temporal extent $t$, and marked initial and final boundaries. When $g,\,x,\,y$ lie inside the joint region of convergence, $\Gamma$, of the series \eqref{eqn:1} for $G(g,x,y,\xi;t)$,  graphs with large area (number of triangles), and long boundaries (number of edges), are exponentially suppressed. Only when the parameters lie on the boundary of $\Gamma$ do large graphs contribute to the sum and therefore the continuum cylinder amplitude is constructed by generalising the scaling limit \eqref{eqn:TwoPtFnScaling} for the two point function $\mathbb G$.
The scaling limit of $\tilde{G}^s$ is defined as (from now on we will suppress the $\xi$ dependence in the amplitudes, but it remains implicitly through the values of $x_c, y_c, f^i_c, g_c$ etc)
\begin{equation}
    \tilde{G}^s(\Lambda,L,L';T)=\lim_{a \to 0} 
    \tilde{G}(g_c-\Lambda a^{d_H}, L/a^{\omega}, L'/a^{\omega};T/a), \label{eqn:Gtildescaling}
\end{equation} where $\omega$ is determined as follows. Note that, by \eqref{eqn:Gtilde},
\begin{equation}
     \tilde{G}^s(\Lambda,L,L';T) = \lim_{a \to 0} \oint_\gamma\frac{du}{2\pi i u}\oint_\gamma\frac{dv}{2\pi iv}\, u^{ L/a^{\omega}}\,v^{ L'/a^{\omega}}\,G(g_c-\Lambda a^{d_H},x_c/u,y_c/v;T/a).
\end{equation}
Changing variables to $X,Y$ defined by 
\begin{eqnarray}
u=1+a^\omega X,\qquad v=&1+a^\omega Y,
\end{eqnarray}
we obtain
\begin{equation}
     \tilde{G}^s(\Lambda,L,L';T) =  \int_{\gamma'}\frac{dX}{2\pi i} \int_{\gamma'}\frac{dY}{2\pi i}\, e^{XL} e^{YL'} G^s(\Lambda,X,Y;T),
\end{equation}
where $\gamma'$ is a path
from $-i\infty$ to $+i\infty$ running to the right of all singularities of the integrand, and
\begin{equation}
      G^s(\Lambda,X,Y;T)= \lim_{a \to 0} a^{2\omega}  G(g_c-\Lambda a^{d_H},x_c(1-a^\omega X),y_c(1-a^\omega Y);T/a).
\end{equation}
For this limit to exist the denominator in \eqref{eqn:amp} must satisfy
\begin{equation}
    1-gx(\xi_2f^2_{t}(y)+(1+gx\xi_3)f^1_{t}(y))=O(a^\omega),\label{denom}
\end{equation}
when expanded about $a=0$ (we have assumed that the numerator in \eqref{eqn:amp} is $O(1)$ which turns out to be the case). At infinite time $f^i_t(y)$ will converge to the fixed point of the evolution equations 
\eqref{eqn:phinfg}, \eqref{eqn:phinftc}, \eqref{eqn:phinfdd}, so 
\begin{equation}
    f^i_t\to f^i_\infty(g,\xi)=f^i_c(\xi)-(\Lambda a^{d_H})^\alpha\phi_c   u^i_{1c} -(\Lambda a^{d_H})^{2\alpha}\chi_c  u^i_{2c}+h.o.t.
\end{equation}
However at large times $f^i_t(y)$ will retain memory of the initial conditions \eqref{eqn:initialconditions} so we expect that
\begin{equation}
    f^i_{T/a}=f^i_c(\xi)+a^{\alpha\, d_H} F^i(\Lambda,Y;T)+h.o.t., \label{fexp11}
\end{equation} 
where $F^i(\Lambda,Y;\infty)=-\Lambda^\alpha\phi_c  u^i_{1c}$ (we will calculate $F^i$ below).
 Vanishing of the $O(1)$ coefficient in \eqref{denom} determines $x_c$, and requiring non-trivial $L$ dependence of $\tilde{G}^s(\Lambda,L,L';T)$ fixes $\omega=d_H\alpha$. 
 
Using \eqref{fexp11} in \eqref{eqn:amp} we see that
\begin{equation}
    G^s(\Lambda,X,Y;T)=-\frac{\partial}{\partial Y}\frac{c_0}{X-c_i F^i(\Lambda,Y;T)},\label{eqn:Gfinalform}
\end{equation}
where $c_0>0, c_{1,2}$ are constants. The amplitude with unmarked final boundary is \begin{equation}
    \tilde{G}_U^s(\Lambda,L,L';T)=\tilde{G}^s(\Lambda,L,L';T)/L',
\end{equation}
so, defining $F(.)\equiv c_i F^i(.)$, 
\begin{align}
    G^{s}_U(\Lambda,X,Y;T)&=\int_{Y}^{\infty} G^s(\Lambda,X,Z;T)\,dZ \label{unmarked1} \\
    &=\frac{c_0}{X-F(\Lambda,Y;T)}-\frac{c_0}{X-F(\Lambda,\infty;T)}. \label{unmarked}
\end{align}
In the pure CDT model \cite{Ambjorn:1998xu}, where  $\xi_i=0, \forall i$ and $c_0=1$, 
$ \tilde{G}_U^s$ satisfies the composition law
\begin{equation}
    \tilde{G}^s_U(\Lambda,L,L';T+T')=\int_{0}^{\infty}\tilde{G}^s_U(\Lambda,L,I;T)\,\tilde{G}^s_U(\Lambda,I,L';T')\,dI, 
\end{equation}
or, in terms of the Laplace transformed quantities, 
\begin{align}
    G^{s}_U(\Lambda,X,Y;T+T')&= \int_{\gamma'}\frac{dZ}{2\pi i} \, G^{s}_U(\Lambda,X,Z;T)\, G^{s}_U(\Lambda,-Z,Y;T')\nonumber\\
    &=\frac{1}{X-F(\Lambda,-F(\Lambda,Y;T');T)}-\frac{1}{X-F(\Lambda,-F_\infty(\Lambda;T');T)},
\end{align}
where $F_\infty(\Lambda;T)= F(\Lambda,\infty;T)$. It follows  that
\begin{eqnarray}
  F(\Lambda,-F(\Lambda,Y;T');T)=F(\Lambda,Y;T+T'),\label{eqn:compoproperty}
\end{eqnarray}
and, taking $T'\to 0$, that $F(\Lambda,Y;0)=-Y$. When type 1 dimers are present this law does not have to be satisfied because the dimer configuration  on the intermediate boundary at $T$ must be summed over as well. However, as we will see below, in some phases the property  \eqref{eqn:compoproperty} \emph{is} still satisfied and then
\begin{eqnarray}
    G^{s}_U&(\Lambda,X,Y;T+T')=c_0^{-1} \int_{\gamma'}\frac{dZ}{2\pi i} \, G^{s}_U(\Lambda,X,Z;T)\, G^{s}_U(\Lambda,-Z,Y;T').\label{eqn:composition}\end{eqnarray}
In these phases the intermediate dimer configuration sum simply generates the $c_0^{-1}$ prefactor.

We will also compute
\begin{equation}
    W(\Lambda,Y)=\int_0^\infty - \frac{\partial F(\Lambda,Y,T)}{\partial Y}\,dT,
\end{equation}
which is the CDT analogue of the disk amplitude for planar random graph models.

\subsection{The differential equations}

To find the differential equations determining $F^i(\Lambda,Y;T)$ we generalize \eqref{eqn:master} away from $t=\infty$  by setting
\begin{equation}
    f^i_{\frac{T}{a}}=f^i_{c}+\phi^i(T),
\end{equation}
and expanding \eqref{eqn:scale1} and \eqref{eqn:scale2} in powers of $a$
which gives
\begin{eqnarray} (1-{\mathbb{T}})^{ij}\phi^j&=&-a\frac{d\phi^i}{dT} -\Lambda a^{d_H}\left(\frac{\partial{{\cal F}^i}}{\partial g}+\frac{\partial {\mathbb{T}}^{ij}}{\partial g}\phi^j\right)
+\frac{1}{2}{\cal F}^i_{\ell k}\, \phi^\ell\phi^k\nonumber\\ &&+\frac{1}{3!}{\cal F}^i_{k\ell m}\phi^k\phi^\ell\phi^m +O\left(a^{2+\alpha d_H},a^{d_H+2\alpha d_H},a^{4\alpha d_H}\right).\label{eqn:timemaster}
\end{eqnarray}
As before,
we decompose $\phi_i(T)$ in terms of $u_{1c}$, $u_{2c}$,
\begin{equation}
    \phi^i(T)=a^{\alpha d_H}\phi_c\Lambda^\alpha\rho(T) u_{1c\,i}+a^{2\alpha d_H}\chi_c\Lambda^{2\alpha}\sigma(T)u_{2c\,i}+h.o.t.
\end{equation}
where by construction $\rho(\infty)=-1$, and $F^i(\Lambda,Y;T)=\Lambda^\alpha\phi_c  u^i_{1c}\,\rho(T)$.
The terms in \eqref{eqn:master} of leading and sub-leading order in $a$, lead, after eliminating $\sigma(T)$, to a differential equation for $\rho(T)$; the constants of integration are fixed using the initial conditions  \eqref{eqn:initialconditions}, and provide the $Y$ dependence of $F^i$. The conditions for the various phases cause some coefficients in \eqref{eqn:timemaster} to vanish which in turn yields different behaviour for $\rho$. We find (in each case $b>0$ is a different $\xi$-dependent constant):
\begin{enumerate}
    \item  Pure-gravity 
\begin{equation}
\frac{1}{b\Lambda^\frac{1}{2}}\frac{\partial \rho}{\partial T}-\rho^2+1=0;\label{eqn:pgde}
\end{equation} 
\item Tri-critical II 
\begin{equation}
\frac{1}{b\Lambda^\frac{2}{3}}\frac{\partial \rho}{\partial T}+\rho^3+1=0; \label{eqn:tctwode}
\end{equation} 
\item Dense Dimer II 
\begin{equation}
\frac{1}{b\Lambda}\frac{\partial \rho}{\partial T}+\rho^3-\rho=0;\label{eqn:ddtwode}
\end{equation} 
\item Tri-critical I 
\begin{equation}
\frac{1}{b^2\Lambda^\frac{2}{3}}\frac{\partial^2 \rho}{\partial T^2} -\frac{3}{b\Lambda^\frac{1}{3}}\rho \frac{\partial \rho}{\partial T} +1 + \rho^3=0;\label{eqn:tconede}
\end{equation} 
\item Dense-dimer I 
\begin{equation}
\frac{1}{b^2\Lambda}\frac{\partial^2 \rho}{\partial T^2} -\frac{3}{b\Lambda^\frac{1}{2}}\rho \frac{\partial \rho}{\partial T} -\rho + \rho^3=0.\label{eqn:ddonede}
\end{equation} 
\end{enumerate}

\subsection{PG, TCII and DDII phases}\label{sec:tc2}

In these phases, only one eigenvalue of $\mathbb T$ reaches 1 at the critical point so $\rho(T)$ is determined by a first order differential equation, and the two initial conditions \eqref{eqn:initialconditions} are in fact equivalent. 

In the PG phase, solving \eqref{eqn:pgde}, we find
\begin{equation}
   F(\Lambda,Y;T)\equiv C\Lambda^\half \rho(T)=C\Lambda^\half\frac{C\Lambda^\half-Y-(C\Lambda^\half+Y) \exp{2b\Lambda^\half T}}{C\Lambda^\half-Y+(C\Lambda^\half+Y)\exp{2b\Lambda^\half T}},\label{eqn:pgfsol}
\end{equation}
where $C=\phi_c  c_i u^i_{1c}$. Re-defining $C^2\Lambda\to\Lambda$, $bC^{-1}\to b$, this leads to the disk amplitude
\begin{equation}
    W(\Lambda,Y)=\frac{1}{\Lambda^\frac{1}{2} +Y},\label{eqn:WPG}
\end{equation}
and the cylinder amplitude
\begin{equation}
    G^s(\Lambda,X,Y;T)=\frac{c_0\Lambda}{\left((XY+\Lambda)\sinh{b\Lambda^\half T}+(X+Y)\Lambda^\half \cosh{b\Lambda^\half T}\right)^2},   \label{eqn:Gpg}
\end{equation}
which are the well known amplitudes for the pure CDT model with no other degrees of freedom \cite{Ambjorn:1998xu} and  apply throughout the pure-gravity phase of the present model. It is straightforward to check that $F(\Lambda,Y;T)$ implies the composition property \eqref{eqn:compoproperty} and therefore that $G^s$ satisfies the composition law \eqref{eqn:composition}. This reflects the fact that the dimers, although present unless $\xi_i=0, \forall i$ (where the discrete model can be solved exactly), are non-critical and generate no long distance correlations in the theory. 

In the TCII phase, integrating \eqref{eqn:tctwode}, applying the initial conditions and re-defining $C^3\Lambda\to\Lambda$, $bC^{-1}\to b$, gives
\begin{equation}
   \exp{-3b\Lambda^\frac{2}{3}T} =\prod_{i=1}^3 \left(\frac{ F(\Lambda,Y;T)+\beta_i \Lambda^\third}{-Y+\beta_i \Lambda^\third}\right)^{\beta_i},\label{eqn:tc2sol}
\end{equation}
where $\beta=(1,e^{i\frac{2\pi}{3}},e^{-i\frac{2\pi}{3}})$ are the cube roots of 1. 
While this relation cannot be solved in terms of elementary functions, it is easy to compute the disk amplitude
\begin{equation}
    W(\Lambda,Y)=\frac{1}{3\Lambda^\third}\frac{2\Lambda^\third+Y}{\Lambda^\frac{2}{3}+Y\Lambda^\third+Y^2},
\end{equation}
and to check that %$\rho(T)$ 
$ F(\Lambda,Y;T)$ still satisfies the composition property (small and large $T$ expansions are given in \ref{app:two}).
 In this phase $\omega=d_H\alpha= \half$ so a typical boundary length $L$ scales as $T^\half$ and the effect of the dimers is to compress the spatial direction rather than induce long range correlations in the temporal direction. An alternative way of seeing this is to observe that the model 
 $ \xi_{1,2,3}=0$, $\xi_4=-1/3$,
lies in this phase in which case there is no dimer configuration to take account of on the intermediate boundary in the composition law; however, the type 4 dimers correlate spatially adjacent vertices leading to the change in scaling behaviour from pure gravity. The tri-critical point of the `natural' model with all $\xi_i$ equal also lies in the TCII phase which, because of the form of $F$, cannot have $G^s$ invariant under the inversion $I$; this cylinder function therefore cannot be representative of the tri-critical phase of the full unrestricted dimer/CDT model for which $G^s$ with all $\xi_i$ equal must be $I$ invariant.

In the DDII phase, integrating \eqref{eqn:ddtwode}, applying the initial conditions and re-defining $C^2\Lambda\to\Lambda$, $bC^{-1}\to b$, gives
\begin{equation}
    F(\Lambda,Y;T) =\Lambda^\half\frac{-Y}{\sqrt{Y^2+(\Lambda-Y^2) \exp{-b\Lambda T}}}, \label{eqn:DD2}
\end{equation}
from which
\begin{equation}
    W(\Lambda,Y)=\frac{2}{Y(\Lambda^\half+Y)}.
\end{equation}
It is again straightforward to check that $F(\Lambda,Y;T)$ satisfies the composition property \eqref{eqn:compoproperty} and therefore that $G^s$ satisfies the composition law. In common with TCII this phase has $\omega=d_H\alpha= \half$ so a typical boundary length $L$ scales as $T^\half$ but the dimer density is large   and again the effect of the  dimers is to compress the spatial direction rather than induce long range correlations in the temporal direction.

\subsection{The TCI phase\label{sec:Scaling limit 3}}
In these phases, both eigenvalues of $\mathbb T$ reach 1 at the critical point so $\rho(T)$ is determined by a second order differential equation, and there are two constants of integration to be determined from the initial conditions \eqref{eqn:initialconditions}. 

The solution for TCI takes the form
\begin{equation}
    F(\Lambda,Y;T)\equiv C\Lambda^\third \rho(T) =-C\Lambda^\third\frac{\sum_{i=1}^3 c_i\beta_i\exp{\beta_i b\Lambda^\third T}}{\sum_{i=1}^3 c_i\exp{\beta_i b\Lambda^\third T}}, \label{eqn:tconesolA}
\end{equation}
where, to ensure that $F$ is real and has the correct behaviour as $T\to\infty$, we must have $c_i\ne0$ and $c_2=c_3^*$. Imposing the boundary conditions shows that $F$ must diverge at $T=0$, which implies that $\sum_ic_i=0$, and that the $O(T^0)$ term is proportional to $Y$. After rescaling $\Lambda$ and $T$ by constants we obtain 
\begin{equation}
    F(\Lambda,Y;T) =-\Lambda^\third\,\frac{\sum_{i=1}^3 (A_c\beta_i Y+\beta_i^2\Lambda^\third)\beta_i\exp{\beta_i \Lambda^\third T}}{\sum_{i=1}^3 (A_c\beta_i Y+\beta_i^2\Lambda^\third)\exp{\beta_i \Lambda^\third T}}, \label{eqn:tconesol}
\end{equation}
and
\begin{equation}
    W(\Lambda,Y)=\frac{1}{A_c^{-1}\Lambda^\frac{1}{3} +Y},
\end{equation}
where $A_c$ is a $\xi$-dependent constant encoding the asymmetry under the inversion operation. We can also solve the discrete equations for TCI exactly leading to the same conclusion, see \ref{app:three}.  Note that $F(\Lambda,Y;T)$ does not have the composition property, and that the same result for $W(\Lambda,Y)$ has also been obtained using the classical limit of a multicritical matrix model \cite{Ambjorn:2012zx}.

Using \eqref{eqn:Gfinalform}, we find the cylinder function
\begin{align}
    G^s(\Lambda,X,Y;T)&=\frac{c_0 A_c F_\infty'(\Lambda;T)}{\left(\left(X-F_\infty(\Lambda;T)\right)\left(A_c Y-F_\infty(\Lambda;T\right))-F_\infty'(\Lambda;T)\right)^2}\label{eqn:tc1G},
\end{align}
where prime is differentiation w.r.t $T$.
Performing the inverse Laplace transform
\begin{align}
    \tilde{G}^s(\Lambda,L_1,A_c L_2;T)=& \nonumber\\
    c_0\exp((L_1+L_2)&F_\infty(\Lambda;T))\,\sqrt{L_1L_2F_\infty'(\Lambda;T)}\,I_1\left(2\sqrt{L_1L_2F_\infty'(\Lambda;T)}\right), \label{eqn:TCIfinaleq}
\end{align}
where $I_1(x)$ is the modified Bessel function of the first kind. The r.h.s. is symmetric between $L_1$ and $L_2$; so $ \tilde{G}^s $ is invariant under the inversion operation for dimer fugacities such that  $A_c=1$.

\subsubsection{Asymptotic Behaviour}
Although the form of \eqref{eqn:TCIfinaleq} applies to both PG and TCI phases, the physical behaviour is quite different. Taking  $T\to0$ for PG gives
\begin{equation}\tilde G^s(\Lambda,L_1,L_2;0)=L_1 \delta(L_1-L_2),
\end{equation}
which is a consequence of the composition law. For TCI the pole in $F_\infty$ leads to the small $T$ behaviour
\begin{equation}
    \tilde{G}_s \sim \exp\left(-2T^{-1}(L_1+L_2-\sqrt{2L_1L_2})\right).\label{eqn:TCIsmallT}
\end{equation}
With $L_1$ fixed, the argument of the exponential above is maximum at $L_2=L_1/2$ so, for small $T$, the length of the final boundary is forced to be half the length of the initial boundary. This slightly non-intuitive behaviour is driven by the negative dimer weights and demonstrates how the TCI phase does not satisfy the composition law.

At large $T$ for TCI,
\begin{align}
    \tilde{G}^s\sim {3\Lambda^{2/3}L_1L_2}e^{-3\Lambda^{1/3}T/2}&\exp\left(-\Lambda^{1/3}(L_1+L_2)\right)\nonumber\\
    &\times\left(-\cos (\sqrt3 \Lambda^{1/3}T/2) +\sqrt3\sin (\sqrt3 \Lambda^{1/3}T/2 )\right).
\end{align}
This has the same functional form of dependence on $L_1,L_2$ but oscillates at large $T$ whereas for PG it decays exponentially without oscillation and is always positive \cite{Ambjorn:1998xu}.
 The average length of the final boundary of TCI universes of temporal extent $T$ that started with length $L_1$ can be 
defined as % 
(\cite{Ambjorn:1998xu}),
\begin{equation}
    \langle L_2\rangle_{L_1,T}=\frac{\int_{0}^{\infty}  G(\Lambda,L_1,L,T)\,L dL}{\int_{0}^{\infty} G(\Lambda,L_1,L,T)\, dL }.
\end{equation}
For small $T$,
\begin{equation}
   \langle L_2\rangle_{L_1,T}= L_1/2 +T+h.o.t,
\end{equation}
which is consistent with the discussion above.
For large $T$, 
\begin{align}    \langle L_2\rangle_{L_1,T}=&\frac{3}{2}\Lambda^{-1/3}+\frac{3}{4}e^{-3\Lambda^{1/3}T/2}\times\nonumber\\&\left(\left(3L_1-4\Lambda^{-1/3}\right)\sqrt{3}\sin(\frac{\sqrt 3 \Lambda^{1/3}T}{2})-3L_1\cos(\frac{\sqrt 3 \Lambda^{1/3}T}{2})\right)+h.o.t.
\end{align}
So for late times it oscillates as it approaches $\frac{3}{2}\Lambda^{-1/3}$ exponentially whereas the pure gravity case approaches a constant value exponentially without oscillations.

\subsection{The DDI phase}

The solution for DDI takes the form
\begin{equation}
    F(\Lambda,Y;T)\equiv C\Lambda^\half \rho(T) =-C\Lambda^\half\frac{c_+ \exp{ b\Lambda^\half T}-c_-\exp{ -b\Lambda^\half T}}{1+c_+ \exp{ b\Lambda^\half T}+c_-\exp{ -b\Lambda^\half T}}. \label{eqn:ddonesol}
\end{equation}
Similar considerations to TCI lead to the disk amplitude 
\begin{equation}
    W(\Lambda,Y)=\frac{1}{\Lambda^\frac{1}{2} +Y},
\end{equation}
and the cylinder amplitude
\begin{align}
    \tilde G^s(\Lambda,L_1,L_2;T)&=\sqrt{\Lambda L_1 L_2/2}\csch(\sqrt \Lambda T)\nonumber\\
    \times &\exp{\left(-\sqrt \Lambda (L_1+L_2)\coth({\sqrt \Lambda T})\right) } \,I_1\left(\sqrt{2\Lambda L_1 L_2}\csch(\sqrt{\Lambda}T)\right). \label{DD1}
\end{align}
These are almost the same as  in the PG phase -- the only difference is that the argument of $I_1$ is $\sqrt 2$ times greater in PG (see \cite{Ambjorn:1998xu} for comparison). So the small $T$ behaviour is similar to TCI \eqref{eqn:TCIsmallT} and no amount of rescaling in \eqref{DD1} can bring it into PG form. This in particular means that, again, the  composition law does not hold.

\section{Time evolution operators \label{Hamiltonian}}

In some phases we can derive the (Euclidean) time translation operator acting on a boundary state space labelled by the length $L$.
From \eqref{eqn:Gfinalform} it follows that
\begin{align}
    \partial_T \, G^s(\Lambda,X,Y;T)=&-\partial_Y\left( \frac{c_0\,\partial_T F(\Lambda,Y;T)}{(X- F(\Lambda,Y;T))^2}\right)\nonumber\\
    =&\,\partial_Y\left(     \frac{\partial_T F(\Lambda,Y;T)}{\partial_Y F(\Lambda,Y;T)} G^s(\Lambda,X,Y;T)\right).
\end{align}
   Assuming the composition property \eqref{eqn:compoproperty} of $F$ we obtain
\begin{equation}
     \frac{\partial_T F(\Lambda,Y;T)}{\partial_Y F(\Lambda,Y;T)}=-\partial_T F(\Lambda,Y;T)\vert_{T=0}.\label{eqn:hreln}
\end{equation}
Evaluating this using \eqref{eqn:pgfsol}, \eqref{eqn:tc2sol}, \eqref{eqn:DD2} gives
\begin{align}
    \partial_T \, G^s(\Lambda,X,Y;T)&=\partial_{Y}\left(b(\Lambda-Y^{2})\,G^s(\Lambda,X,Y;T\right),\\
     \partial_T \, G^s(\Lambda,X,Y;T)&=\partial_{Y}\left(b(\Lambda-Y^{3})\,G^s(\Lambda,X,Y;T)\right),\\
      \partial_T G^s(\Lambda,X,Y;T)&=\partial_{Y}\left(b(\Lambda Y-Y^3)\,G^s(\Lambda,X,Y;T)\right),
\end{align}
for pure gravity,
 TCII, and DDII respectively. Taking Laplace transforms shows that the doubly marked amplitudes satisfy the evolution equation 
 \begin{equation}
     -\partial_{bT}\,\psi(L) = H(L,\partial_L)\,\psi(L),
 \end{equation}
 with 
 \begin{align}
     H_{PG}=& -L\partial_L^2+L\Lambda,\label{eqn:HPG}\\
      H_{TCII}=& -L\partial_L^3+L\Lambda,\\
       H_{DDII}=& -L\partial_L^3+L\Lambda\partial_L.
 \end{align}
Only $H_{PG}$ \cite{Ambjorn:1998xu} is self-adjoint and can be regarded as a Hamiltonian acting on the $L$ basis. $H_{TCII}$ was also obtained in \cite{Atkin:2012yt} for the classical limit of a multicritical matrix model and is non-self-adjoint. Curiously, $H_{DDII}$ is anti-self-adjoint. For TCI and DDI the right hand side of \eqref{eqn:hreln} diverges which reflects the fact that the composition property does not hold and the sum over the intermediate dimer configuration is non-trivial. In fact one could take the converse point of view; namely that, because we expect the critical dimers to induce correlations in the $T$ direction, 
$\partial_T F$ \emph{must} diverge at $T=0$.
Of course the left hand side of \eqref{eqn:hreln} exists and is finite for $T>0$ but it is $T$ dependent so there is no time-independent evolution operator for these phases.

\section{Discussion}\label{sec:discussion}
In this paper we extended previous solutions of the hard dimer model on causal triangulations to allow all types of dimer on the graphs, and by removing all but one of the constraints imposed.
The only constraint remaining in this paper is, in the language of labelled trees, that it is forbidden for a last child to take label $p=3$ (see \ref{itm:3left} in Section \ref{sec:Bijection to Trees}). This constraint is sufficient to  render the tree model local, in the sense that there are no correlations between vertices at the same height that are not siblings, and hence exactly solvable. The model has a number of phases, depending upon the dimer weights $\xi$, which are all controlled by the quadratic and cubic critical points of a polynomial. The remaining question is whether our calculation is sufficiently general that, even if the last constraint were removed, no further critical point with distinct properties would appear.

The TCII and DDII phases obey the naive composition law because the dimers are correlated in the spatial direction but not strongly so in the $T$ direction. As the constraint affects spatially neighbouring dimers only it seems unlikely  that its removal will cause new correlations in the $T$ direction to develop; indeed  our solution is exact for the $ \xi_{1,2,3}=0,\xi_4=-1/3,$ TCII model which is unaffected by the constraint. So we expect that the TCII phase will survive for some $\xi$, but that its boundaries will move; the same applies to the DDII phase as it appears on the boundary of TCII. However,  %$F(\Lambda,Y;T)$
$G^s(\Lambda,X,Y;T)$ in these phases 
can never be inversion symmetric because of \eqref{eqn:tc2sol} and \eqref{eqn:DD2}; if the $\xi_3=\xi_4$ tri-critical points lie in TCII in   the unconstrained model, then $G^s(\Lambda,X,Y;T)$ must be modified. 

The TCI and DDI phases have strong dimer-induced correlations in both temporal and spatial directions. This is reflected in the appearance of the second unit eigenvalue at criticality, isotropic scaling, and a cylinder function that does not obey the naive composition law.  
Unless the second eigenvalue is destroyed, these phases should be robust against removal of the constraint. When $\xi_3=\xi_4$, the full unconstrained model satisfies the inversion symmetry so $G^s(\Lambda,X,Y;T)=G^s(\Lambda,Y,X;T)$ which is satisfied by \eqref{eqn:tc1G} whenever the asymmetry parameter $A_c=1$. Provided a completely new phase does not appear we expect that TCI does characterize the tri-critical physics of the full unconstrained but inversion-symmetric model.

%\textcolor{blue}
{The scaling limit of the pure 2d CDT has been shown \cite{Ambjorn:2013joa} to be two-dimensional projectable Ho\u rava-Lifshitz (HL) gravity \cite{Horava:2009uw,Horava:2010zj}. HL gravity assumes a foliated space-time so the full diffeomorphism group is restricted to to the foliated diffeomorphisms for which $t\to t+\xi^0(t)$, and $x\to x +\xi^1(t,x)$. At least naively one can then regard the time-slicing of CDT as a gauge-fixed discretization of the foliated continuum space. In the projectable  version of HL %Ho\u rava-Lifshitz 
gravity the lapse function (the time-time component of the metric) is assumed to be a function of $t$ only. Then one can compute the scaling Hamiltonian and the disc amplitude analytically and obtain exactly \eqref{eqn:HPG} and \eqref{eqn:WPG} respectively, thus showing that the scaling limit of the pure CDT and HL gravity are indeed equivalent.
On the other hand the critical point of the hard dimer model on a fixed lattice is known to correspond to a conformal field theory with central charge $c=-\frac{22}{5}$ which also describes the Yang-Lee edge singularity \cite{Cardy:1985yy}.
   We therefore expect that each scaling limit found  here corresponds somehow to a CFT coupled to Ho\u rava-Lifshitz gravity, and in particular that one of them is the Yang-Lee singularity -- so which one?}
   
 % \textcolor{orange}
 { The case when a lattice model whose critical dynamics is described by a CFT is coupled to DT is well understood; the matter exponents are shifted according to the KPZ formula \cite{Knizhnik:1988ak}, and the geometry of the gravity sector is changed. 
  For  $c>1$ the interaction between matter and geometry is so strong that the space degenerates into a branched polymer structure. As $c$ is decreased from one, the interaction becomes steadily weaker until at $c=-\infty$ matter exponents are unchanged from fixed lattice values and the space is smooth and flat and two-dimensional. Numerical evidence suggests that the behaviour of CDT is rather different with weaker interaction between matter and geometry, and in particular that $d_H=2$ when $0\le c<1$. At $c=-\infty$ we again expect to find that space is smooth, flat and two-dimensional so it is likely that in fact $d_H=2$ for all $c\le 1$ . The phase with the least back reaction from the  dimers onto the geometry is DDI which indeed has the same $d_H=2$  as the pure gravity phase; the fact that the local Hausdorff dimension $d_h=3$, rather than $d_h=2$, is not observable in the scaling limit (but it is a signature that there is local interaction between dimers and geometry). The  DDI disc amplitude is the same as for  pure gravity, and the cylinder amplitude alone of the geometrical observables betrays the presence of the dimers. As discussed in section  \ref{subsec:phasediag}, the dimer exponent is shifted from its fixed lattice value, $\sigma=-\sixth$, to $\sigma=-\third$ as predicted by the KPZ formula. The fine tuning needed to get this rather than $\sigma=\half$, as seen in TCI, precisely decouples the dimer degrees of freedom from the geometry and turns off the operator mixing discussed by \cite{Ambjorn:2014jca}. These arguments all point to the scaling limit of the DDI phase being the the Yang-Lee edge singularity coupled to CDT.}
  
%\textcolor{purple}
{ Unlike 1+1d pure CDT, but in common with $3+1$ dimensional CDT, the model that we have studied in this paper has several coupling constants whose variation maps out a number of different phases. The DDI and TCI phases are genuinely two dimensional; they have isotropic scaling and non-trivial correlations in the time direction. However in the DDII and TCII phases the scaling behaviour in the spatial and temporal directions is very different, and the dimers do not induce strong correlations in the temporal direction. Simulations of $3+1$ dimensional CDT with toroidal spatial topology have demonstrated the existence of phases in which there is little temporal correlation between spatial slices \cite{Ambjorn:2021yvk}. Thus the existence of asymmetric phases, in addition to phases where the scaling behaviour of the geometry is isotropic, seems to be a characteristic of CDT with some topologies regardless of dimension, and at least in 1+1d can be demonstrated analytically.}

\omittext{\ack PDX acknowledges the support of an Oxford Berman Scholarship.}

\appendix
\section{Proof that \texorpdfstring{$\gamma$}{Lg} is a bijection}\label{bijectionproof}
\begin{proof}
Injectivity follows from checking that the map $\gamma^{-1}$ exists. This is constructed by combining $\beta^{-1}$,  the inverse map $\mathcal{P}_{t+1}\to\mathcal{T}_t$  \cite{Durhuus:2009sm}, with  the labelling rules (\ref{firstrule})-(\ref{lastrule}) followed backwards.
To check surjectivity we show that $\mathcal{RL}_{t+1}$ is indeed the image of $\mathcal{D}_t$ under the map $\gamma$,
%. This means 
by verifying constraints   \ref{itm:initial}-\ref{itm:4leftmost}.

\begin{itemize}
    \item \ref{itm:initial} and \ref{itm:final} follow from the boundary conditions that there are no dimers dual to edges on $\partial_1 T$ and $\partial_2 T$ respectively. 

\item \ref{harddimer} 
follows from the hard dimer rule.
    To deduce the rules consider the local occurrence of any dimer type from Fig. \ref{fig:DimerTypes} corresponding to the label $\ell(v)$ of vertex $v$. Then add a `probe' dimer to the vacant edges in turn, thus violating the dimer rule. Each time attach a new triangle to the same edge in all allowed ways and apply $\gamma$ to map the segment of the triangulation to a segment of a labelled tree; if the probe dimer corresponds to the labelling of a vertex $w$ which is $v$ itself or next to $v$ or a child of $v$, then there is a new constraint on $\ell(w)$, otherwise not. 
    Figs \ref{fig:RestrictionsDueTox=3}, \ref{fig:RestrictionsOnChildrenOfx=2} and \ref{fig:RestrictionsOnNeightboursOfy=4} show how the constraints \ref{itm:3pres}, \ref{itm:2des} and \ref{itm:4pres} respectively arise. \ref{itm:onefour} follows from applying the same procedure, starting with dimer type $1$.     
    \end{itemize}
\begin{figure}[H]
    \centering
    \includegraphics[scale=0.4,trim={0 9cm 0 0},clip]{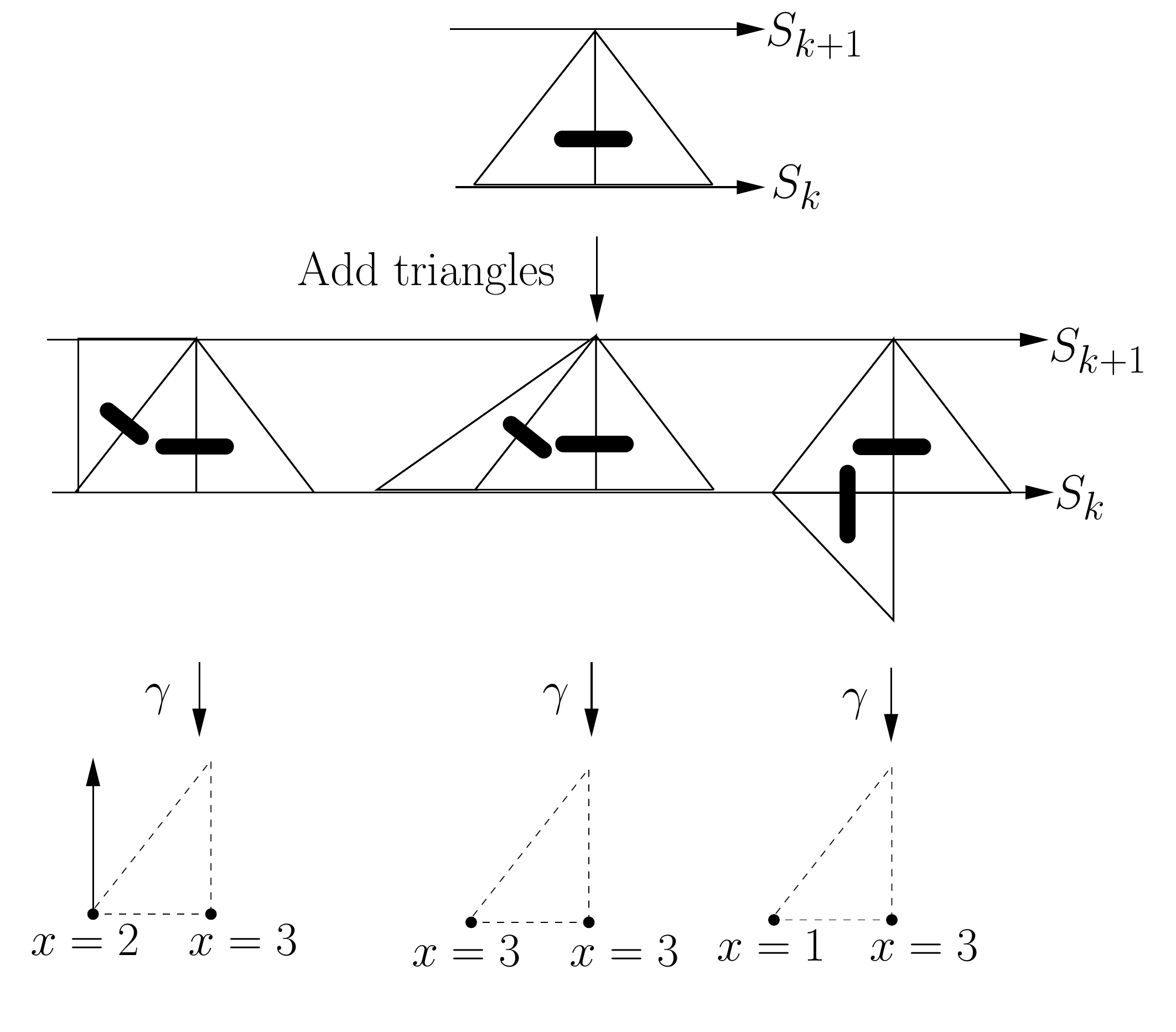}
    \caption{\label{fig:RestrictionsDueTox=3}The dashed lines are the edges of the triangulation while the solid lines are those of the tree. }\end{figure} 
\begin{figure}[H]
    \centering
    \includegraphics[scale=0.4,trim={0 7.7cm 0cm 0},clip]{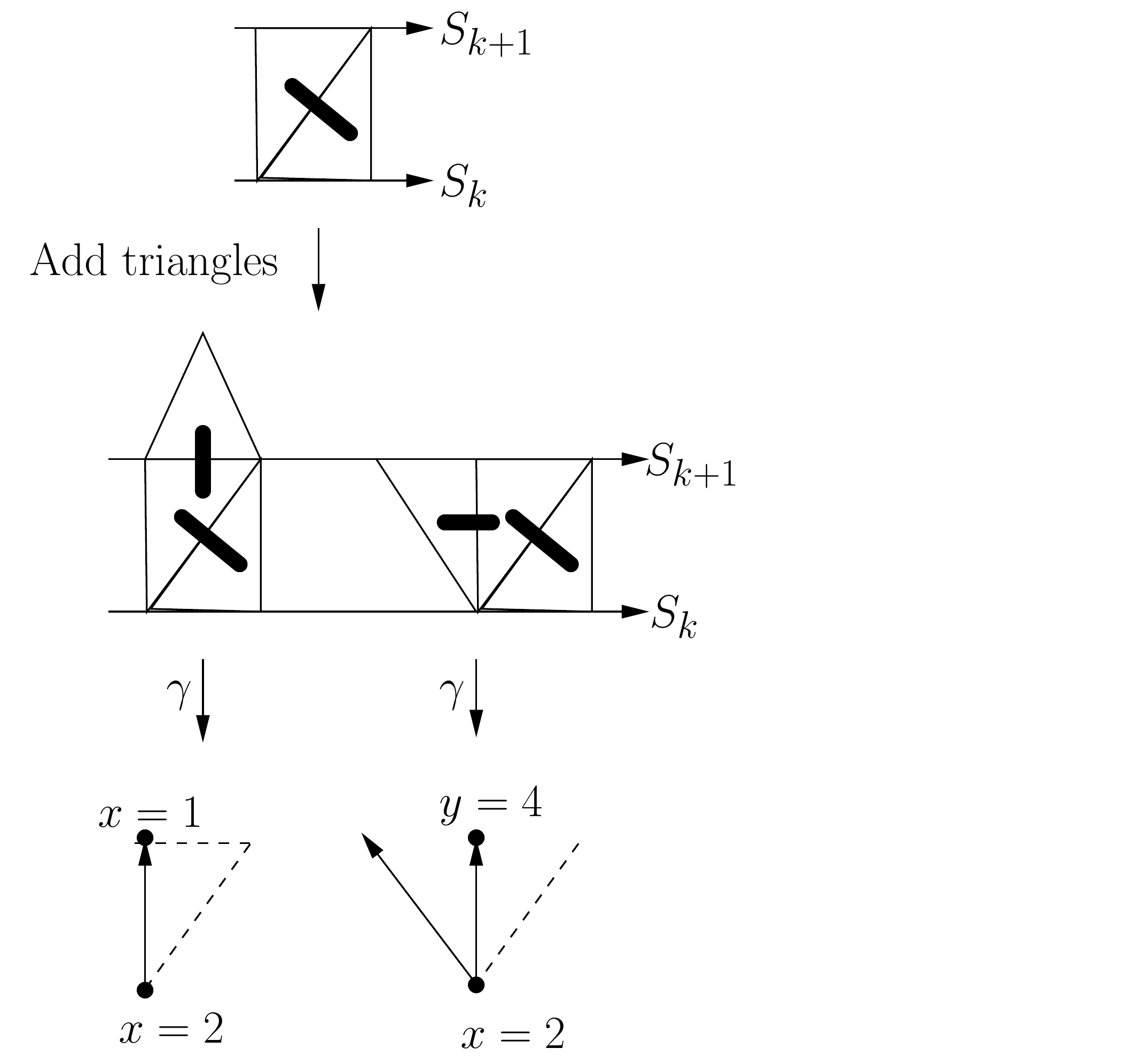}
    \caption{\ref{itm:2des}. \label{fig:RestrictionsOnChildrenOfx=2}}
\end{figure} 
\begin{figure}[H]
    \centering
    \includegraphics[scale=0.4,trim={0 9cm 0cm 0},clip]{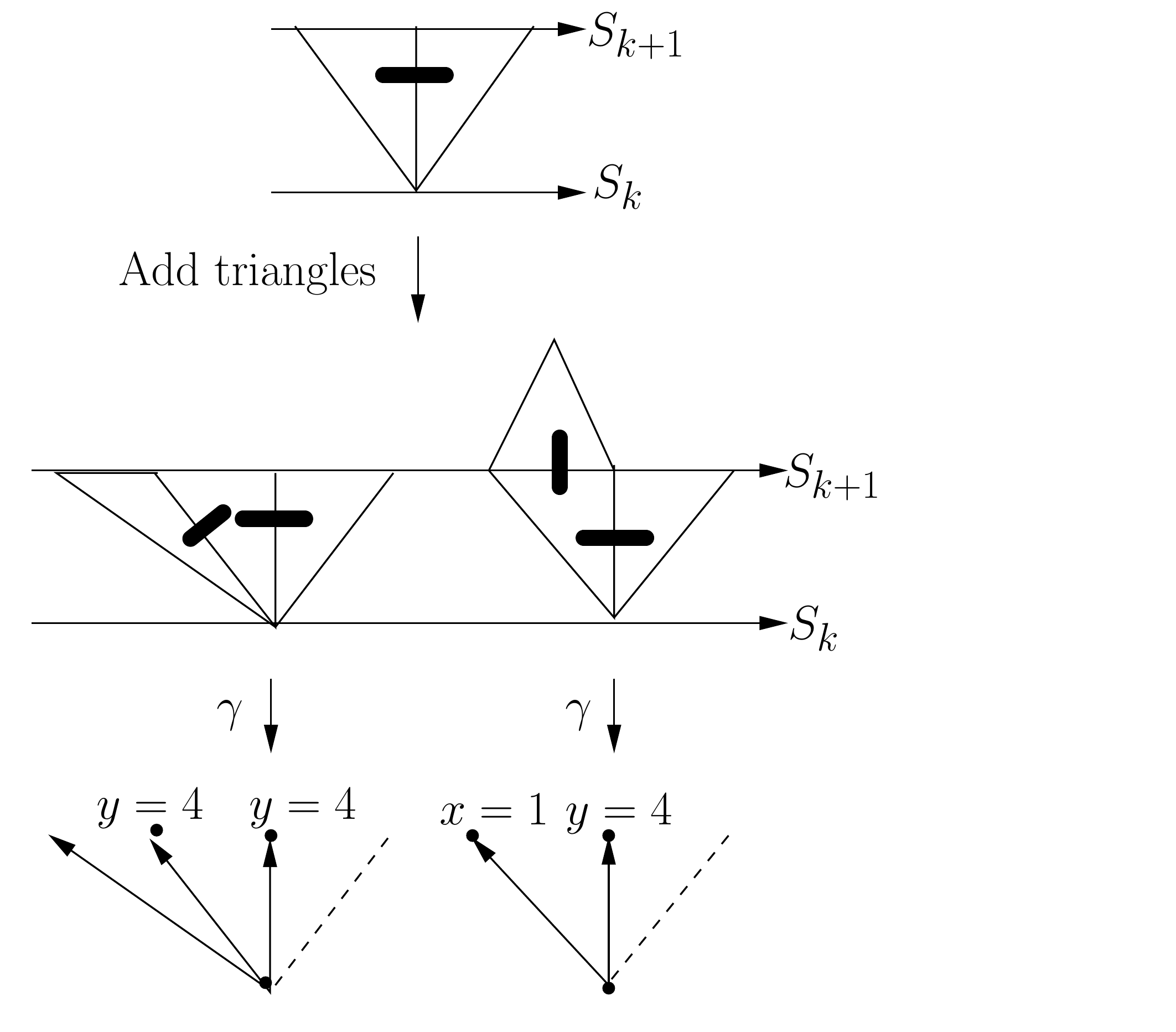}
    \caption{\ref{itm:4pres}. \label{fig:RestrictionsOnNeightboursOfy=4}}
\end{figure} 
\begin{itemize}
   \item \ref{itm:3left} follows from the restriction (see Fig. \ref{fig:restriction1}).
   \end{itemize}
   
   \begin{figure}[H]
    \centering
    \includegraphics[scale=0.4,trim={0 21cm 0 0},clip]{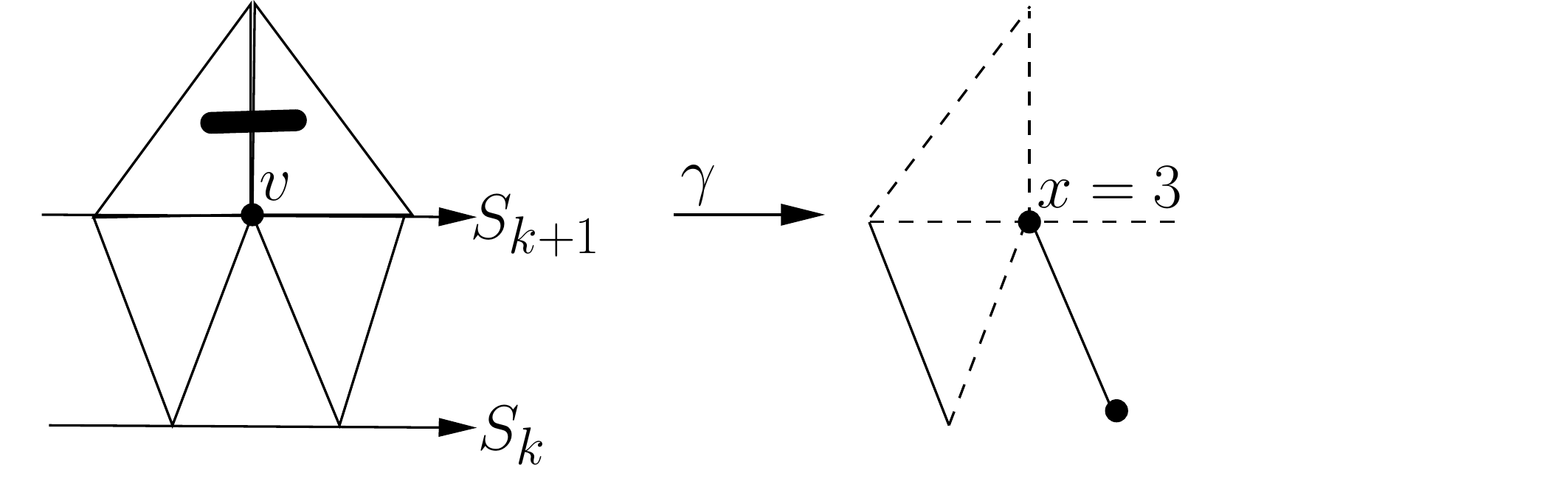}
    \caption{\label{fig:restriction1} We apply $\gamma$ to a triangulation with $f(v)$ dual to a dimer, $\sigma_f(v)=1$ and $\sigma_b(v)=2$ to obtain \ref{itm:3left}.}
\end{figure} 

   \begin{itemize}
    \item \ref{itm:leaf}, \ref{itm:2atleastone} and \ref{itm:4leftmost} follow from the geometry of the dimer types displayed in Fig. \ref{fig:DimerTypes}. Fig. \ref{fig:RestrictionsDueToGeometry} illustrates how one deduces them. 
\end{itemize}
\begin{figure}[H]
    \centering
    \includegraphics[scale=0.4,trim={0 15cm 0 0},clip]{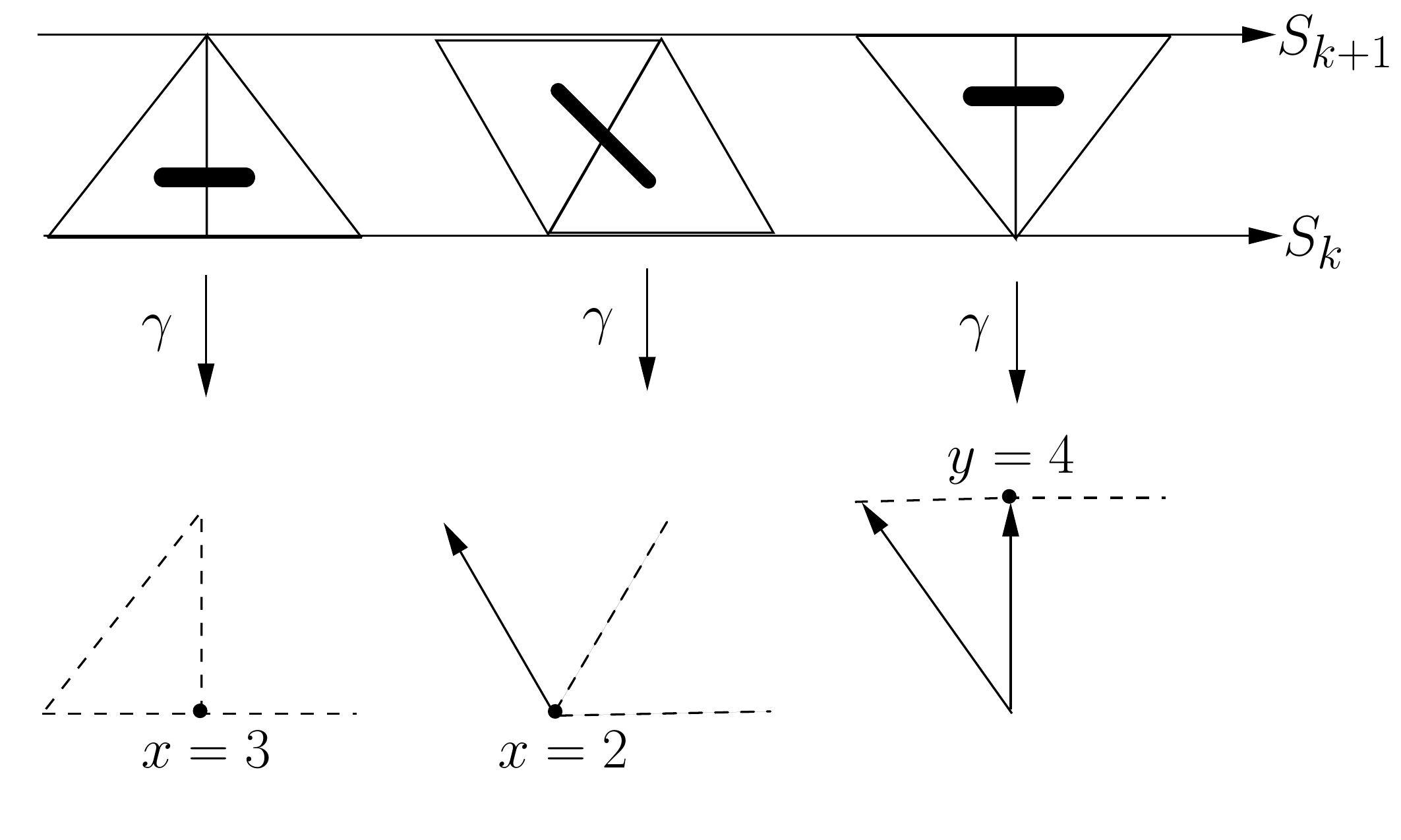}
    \caption{\label{fig:RestrictionsDueToGeometry}We start with the definition of a specific dimer type and study the segment of the triangulation under the map $\gamma$. The dashed lines are the edges of the triangulation while the solid lines are those of the tree.}\end{figure} 
\end{proof}

\section{Cylinder Amplitude Series for TCII}\label{app:two}

For TCII the cylinder amplitude cannot be calculated in closed form so we record the series expansion  at high and low $T$ for completeness. Considering the model 
\begin{equation}
    \xi_1=\xi_2=\xi_3=0,\quad\xi_4=-1/3,
\end{equation}
for convenience, and defining $g=g_c-\Lambda a^3/18+...$, $y=\sqrt{3}-aY+...$, the  small $T$ expansion of  $F(\Lambda,Y;T)$ takes the form
\begin{align}
    F(T,\Lambda,Y)&=-Y+\frac{1}{3} T
   \left(Y^3-\Lambda \right)+\frac{1}{6} T^2 Y^2
   \left(\Lambda
   -Y^3\right)+\frac{1}{54} T^3
   \left(Y^3-\Lambda \right) \left(5
   Y^4-2 \Lambda 
   Y\right) \nonumber \\
   &+\frac{1}{648} T^4 \left(\Lambda
   -Y^3\right) \left(2 \Lambda ^2+35
   Y^6-28 \Lambda 
   Y^3\right)+\ldots . \label{subleading}
\end{align}

For large $T$, $F(T,\Lambda,Y)$ can be expanded as a Taylor series in $\exp(-\Lambda^{2/3}T)$ and we get 
\begin{equation}F(T,\Lambda,Y)=-\Lambda^{1/3}\sum_{n=0}a_n \left(\alpha \phi(Y)\right)^n e^{-n\Lambda^{2/3}T}, 
\end{equation}
where 
\begin{align}
\phi(Y)&=\prod_i(-Y+\alpha_i)^{\alpha_i},\nonumber\\
\alpha&=-(-1+\alpha_2)^{-\alpha_2}(-1+\alpha_3)^{-\alpha_3},
\end{align}
and the first few coefficients are $a_0=1, a_1=1, a_2=1, a_3= 7/6, a_4=13/9, a_5=133/72, a_6=217/90$.

\section{Discrete solution for TCI Cylinder Amplitude}\label{app:three}

On TCI, $\xi_4=0$, so, eliminating $f^2_t$ from \eqref{eqn:scale1} and \eqref{eqn:scale2}, $f^1_t$ satisfies
\begin{align}
    R(f_{t-1},f_t,f_{t+1})\equiv&\nonumber\\
    g^4 \xi _1 \xi _2 f_{t-1} f_t f_{t+1}&-g^2 f_t f_{t+1} \left(g^2 \xi _3+\xi _1+\xi _2+1\right)+f_{t+1} \left(g^2 \xi
   _2+1\right)-1=0, \label{thereccu}
\end{align}
and the corresponding solution is
\begin{equation}
    f_t=\frac{c_1 p_1^{-t}+c_2 p_2^{-t}+c_3 p_3^{-t}}{c_1 p_1^{-t-1}+c_2 p_2^{-t-1}+c_3 p_3^{-t-1}}, \label{solution1}
\end{equation}
where $c_{i}$ are unfixed constants and $p_{i}$ are the three roots of $R(p_i,p_i,p_i)=0$.
For ease of presentation, we do the calculation with the critical couplings $\xi_3=0, \xi_1=\xi_2/4=-1/8$ %. As such,
for which $f^1_c=f^2_c=4$ and $g_c=1/\sqrt{2}$. Then we take $g=g_c(1-\Lambda a^3/4)$.
First, impose the boundary conditions \eqref{eqn:initialconditions}  to determine the constants $c_i$ in terms of the final boundary fugacity $y$. Expanding $p_i$ in $a$ gives $p_i=4(1-\alpha_i\Lambda^{1/3}a)+O(a^2)$, where $\alpha_i$ are the 3 roots of 1. 
Then the numerator and denominator of \eqref{solution1} at leading order read $4^{-t}\sum C_i \exp(\alpha_i \Lambda^{1/3}T)$ and $4^{-t-1}\sum C_i \exp(\alpha_i \Lambda^{1/3}(T+a))$, where $C_i$ is the leading term of $c_i$. With $u(T)=\sum_i C_i \exp(\alpha_i \Lambda^{1/3}T)$, \eqref{solution1} then becomes 
\begin{equation}
    f_{T/a}=4u(T)/u(T+a)=4(1-a\, u'(T)/u(T))+O(a^{2}).
\end{equation} 
Comparing this with \eqref{fexp11} we see that $F=-4u'/u$. 
Calculating $c_i$ using the boundary conditions, one finds that $c_i=\alpha_i+a \frac{(...)}{y_c-1/\sqrt 2}+O(a^2)$, where (...) is a constant. This indicates that the non-trivial scaling occurs  at $y_c=1/\sqrt 2$. Setting $y=(1-aY)/\sqrt2$, we get $c_i=\alpha_i(2Y+\Lambda^{1/3}\alpha_i)+O(a)$, i.e $C=\alpha_i(2Y+\Lambda^{1/3}\alpha_i)$. Therefore
\begin{equation}
    F(T,\Lambda,Y)=-4\sqrt[3]{\Lambda}\frac{\sum_i (\alpha_i^2Y+\sqrt[3]{\Lambda}/2)\exp(\alpha_i\sqrt[3]{\Lambda}T)}{\sum_i (\alpha_i Y+\alpha_i^2\sqrt[3]{\Lambda}/2)\exp(\alpha_i\sqrt[3]{\Lambda}T)}. \label{eqn:f*}
\end{equation}

\section*{References}
\bibliographystyle{JHEP}
\bibliography{HDCDTbiblio} % Entries are in the "HDCDTbiblio.bib" file

\end{document}